%% file: displace_sensor16.tex
\documentclass[aps, pra, reprint, amsmath,amssymb,showpacs]{revtex4-1}

\usepackage{qcircuit}
\usepackage{subfigure}
\usepackage{mathtools}
\usepackage{amssymb}
\usepackage{amsfonts}
\usepackage{amsmath}
\usepackage{amsthm}
\usepackage{dsfont}
\usepackage{latexsym}
\usepackage{dcolumn}

\usepackage{amsmath}
\usepackage{braket}
\usepackage{graphicx}

\newcommand{\appropto}{\mathrel{\vcenter{
  \offinterlineskip\halign{\hfil$##$\cr
    \propto\cr\noalign{\kern2pt}\sim\cr\noalign{\kern-2pt}}}}}

\newcommand{\ketbra}[2]{| #1\rangle\! \langle #2|}
\newcommand{\leftlabel}[1]{\push{\phantom{#1}}& \lstick{#1}}

\usepackage{psfrag,pstricks}

\begin{document}
\title{Single-Mode Displacement Sensor}

\author{Kasper Duivenvoorden, Barbara M. Terhal and Daniel Weigand}
\affiliation{JARA Institute for Quantum Information, RWTH Aachen University, 52056 Aachen, Germany}
\date{\today}

\begin{abstract}
We show that one can determine both parameters of a displacement acting on an oscillator with an accuracy which scales inversely with the square root of the number of photons in the oscillator. Our results are obtained by using a grid state as a sensor state for detecting small translations in phase space (displacements). Grid states were first proposed in \cite{GKP} for encoding a qubit into an oscillator: an efficient preparation protocol of such states, using a coupling to a qubit, was developed in \cite{TW:GKP}.  We compare the performance of the grid state with the quantum compass or cat code state and place our results in the context of the two-parameter quantum Cram\'er-Rao lower bound on the variances of the displacement parameters. We show that the accessible information about the displacement for a grid state increases with the number of photons in the state when we measure and prepare the state using a phase estimation protocol. This is in contrast with the accessible information in the quantum compass state which we show is always upper bounded by a constant, independent of the number of photons.  We present numerical simulations of a phase estimation based preparation protocol of a grid state in the presence of photon loss, nonlinearities and qubit measurement, using no post-selection, showing how the two effective squeezing parameters which characterize the grid state change during the preparation. The idea behind the phase estimation protocol is a simple maximal-information gain strategy.
 \end{abstract}
\maketitle

\section{Introduction}
It is a fact of quantum mechanics that one cannot simultaneously measure the position $q$ and momentum $p$ of a quantum state as the position and momentum operator do not commute $[\hat{q},\hat{p}]=i\hbar$. This however does not imply that every function of these operators cannot be measured simultaneously as some functions of these operators may commute. In this paper we show how one can use this fact to simultaneously learn both parameters of a displacement acting on an oscillator. We assume a harmonic oscillator with annihilation operator $a=\sqrt{\frac{m\omega}{2\hbar}}(\hat{q}+\frac{i\hat{p}}{m\omega})$ with mass $m$, angular frequency $\omega$ and Hamiltonian $H=\hbar \omega (a^{\dagger} a+\frac{1}{2})$.
%DW language
This harmonic oscillator for example could physically represent an optical or microwave cavity mode, an LC oscillator or a nanomechanical oscillator.

% dimension space is \sqrt(hbar/m omega}, dimension momentum is sqrt{m\omega \hbar}, so  S_p=exp(ip/hbar (dimensionfull shift) and S_q=exp(i q/hbar dimensionfull shift)

 Assume that one has prepared a state in the harmonic oscillator at some initial time $t=0$: we will refer to this state as the sensor state. For an interval in time from $t=0$ to $\tau$ a weak time-dependent classical force acts so that during this interval the Hamiltonian equals $H(t)=\hbar\omega (a^{\dagger} a +\frac{1}{2})- \hat{q} F(t)$. The unitary evolution on the oscillator equals (up to a phase) $U(\tau)=\exp(-i \omega a^{\dagger} a \tau) \exp(\beta a^{\dagger}-\beta^* a)$ with $\beta=\frac{-i}{\sqrt{2m \omega \hbar}} \int_0^{\tau} dt' F(t') e^{i \omega t'}$, thus picking up an additional displacement $D(\beta)\equiv \exp(\beta a^{\dagger}-\beta^* a)$. Determining the amplitude and phase of the displacement parameter $\beta$ will roughly provide information about the amplitude and phase of the force at frequency $\omega$. If the harmonic oscillator represents, say, a superconducting LC oscillator, then a capacitive coupling to a time-dependent classical gate voltage can similarly induce small displacements: the goal is then to simultaneously measure both amplitude and phase of the voltage signal \footnote{Since we want to treat the field causing the displacement classically, it should contain a fair number of photons, but the effect on the oscillator should be weak, adding at most $\pi/2$ photons, hence one requires weak coupling and/or the oscillator and the signal being off-resonant.}.
 
  %DW Language
 The harmonic oscillator can also represent a nanomechanical oscillator,  e.g. in \cite{RTL:nanomechanical} the strength of a small displacement on a nanomechanical oscillator is measured via the induced shift of the resonant frequency of a coupled superconducting microwave cavity mode. Determining the ultimate limits of measuring the effect of a weak force on an oscilator has been a subject of much previous study, see e.g. \cite{caves+:rmp}. It is now known that more complex measurement strategies do not necessarily obey `a standard quantum limit' (SQL), see e.g. \cite{GLM:quantum-enhanced}, but can reach a Heisenberg limit.
  For example, in \cite{penasa+:displacement} the amplitude of a small displacement acting on a microwave cavity field was measured using an entangled atom-cavity sensor state, going beyond the SQL, performing best when the displacement direction or phase is known beforehand (see also a Rydberg-atom based electrometer in \cite{facon+:electrometer}).
%DW language
The focus of much theoretical and experimental research in parameter estimation for oscillators has been on measuring a phase shift, of the form $\exp(i \phi a^{\dagger} a)$: this is not the setting that we consider here \footnote{A (dispersive) phaseshift can come about when the linear coupling between oscillator and the external field is off-resonant while a real displacement is induced when the coupling is on-resonant.}. The idea that we put forward in this paper does not directly relate to a discussion on a SQL versus Heisenberg limit: the essential new idea is that we show that one can measure {\em both} parameters in a displacement simultaneously.
 
Before we provide an overview of our results, let us discuss, at a descriptive level, how several quantum states could perform as displacement sensors.
 
A simple sensor state that one can consider for displacement sensing is a coherent state $\ket{\alpha}$ (for example a vacuum state). For such a coherent sensor state it is clear that the shot noise in both quadratures limits the accuracy with which one can determine the parameters in a displacement $D(\beta)$ mapping $\ket{\alpha}$ onto $\ket{\alpha+\beta}$. 

%The optimal measurement for determining the displacement is heterodyne detection, or a measurement in the overcomplete basis of coherent states.
% DW language
Using squeezed states would improve the sensitivity for one of the shift parameters while at the same time losing accuracy in estimating the other parameter, suggesting that for a single-mode sensor it might never be possible to get high accuracy for both displacement parameters.

% Note that the goal is here in displacement sensing is not to transmit information through a channel using coherent or other signal states

An effective way of determining an unknown displacement is to let such a displacement act on only one mode of a two-mode squeezed state as in the superdense coding protocol introduced by Kimble and Braunstein \cite{BK:superdense}. The infinitely-squeezed two-mode squeezed state --the paradigmatic EPR state-- is a simultaneous eigenstate of both the difference of the positions of the oscillators $\hat{q}_1-\hat{q}_2$ and the sum of the momenta $\hat{p}_1+\hat{p}_2$, this being possible because these operators commute. Itinerant (traveling over a transmission line, not confined in a cavity) single and two-mode squeezed states in the microwave domain have been experimentally generated \cite{flurin+:2mode, mallet+:squeezed} (see an overview in \cite{NY:microwave}).

%For finite amounts of squeezing one can describe the performance of such a sensing protocol by the mutual information obtained between the input displacement `signal' $\beta$, described by some prior probability distribution, and the estimated output signal $\tilde{\beta}$. In the superdense coding protocol this information is obtained through homodyne measurements on the two modes and the information that is gained has been shown to scale as $\Theta(\log \overline{n})$ with $\overline{n}$ the average number of photons \cite{BK:superdense}. Unlike the idea of the single-mode displacement sensor based on the grid state, this two-mode displacement sensor is also effective when the displacements are large, i.e. there is no restriction that $|\beta|$ takes values in a small interval. 

Another simple method of measuring a displacement is to have two bosonic probe modes which {\em both undergo the same displacement} $D(\beta)$. One probe mode will be prepared in a squeezed position state, the other probe mode in a squeezed momentum state, so that both shifts can be determined with high accuracy bounded by the amount of squeezing available.   

Another state that has been suggested as a good displacement sensor state is the single-mode quantum compass state \cite{zurek:sub}, also known as a particular cat code state \cite{vlastakis+:cat100, leghtas+:QEC, BMT:review, book:haroche}. It has been argued in \cite{zurek:sub} that this state equal to 
 \begin{equation}
 \ket{\psi^{\rm comp}}\propto \ket{\alpha}+\ket{-\alpha}+\ket{i \alpha}+\ket{-i \alpha} \nonumber,
 \end{equation}
   has a phase space structure in the central interference region (see Fig.~(\ref{fig:wigner})) with interference `tiles' with an area $\sim (2 \pi \hbar)^2/\overline{n}$, thus allowing for a sensitivity which increases with $\overline{n}$. A small, $\overline{n}=|\alpha|^2\approx 3$, version of this state has been recently used to store a qubit in a microwave cavity \cite{ofek+:QEC}. When $\overline{n} \gg |\beta|$ one can show that $\bra{\psi^{\rm comp}} D(\beta) \ket{\psi^{\rm comp}} \propto \cos(\alpha{\rm Im}(\beta))+\cos(\alpha {\rm Re}(\beta))$ so that if ${\rm Im}(\beta)={\rm Re}(\beta)=\pi/(2 \alpha)\sim \pi/\sqrt{\overline{n}}$, i.e. the displaced state is orthogonal to $\ket{\psi^{\rm comp}}$ itself. The state is thus sensitive to very small displacements of strength $|\beta| \sim  1/\sqrt{\overline{n}}$, but that does not imply that it performs well when $|\beta|$ lies in a constant range larger than $1/\sqrt{\overline{n}}$. 

% BMT summary
Hence the goal of this paper is to show that for small displacements the use of two modes (of which either only one or both undergo the unknown displacement) is unnecessary: using a single-mode grid state sensor one can get displacement information about both parameters which also gets better with the number of photons in the state. This result holds for displacements of small strength, adding at most $\frac{\pi}{2}$ photons to a vacuum state.  The results in this paper are focused on the {\em single-shot} setting: that is, preparation of the sensor state, the application of the displacement, and the subsequent measurement of the effect, is done {\em once}, no repetition of the experiment is allowed.
%DW Languaga
In quantum metrology one often considers the setting where an identical experiment (set up of sensor state, displacement happens, measure) can be repeated to gain information. We briefly discuss the repetition set up in the Conclusion. In any scenario, single-shot or repetition, the point of the grid state is that one has effective squeezing in both quadratures thus an enhanced sensitivity in both.

%DW Language
In Section \ref{sec:idea} we give the idea behind using grid states. In Section \ref{sec:estim} we discuss the background on two-parameter estimation theory. In Section \ref{sec:estimphase} we discuss the phase estimation protocols which allow one to prepare a grid state as well as measure the effect of a displacement. We will explicitly prove how the accuracy with which one can determine both parameters in the displacement increases with photon number for grid states which are explicitly prepared and measured using a noiseless textbook phase estimation protocol. In Section \ref{sec:qcompass} we compare our results with a quantum compass state sensor in terms of how much accessible information one can get about a displacement based on the phase estimation measurement.
In Section \ref{sec:exp} we report on extensive numerical simulations of executing a phase estimation protocol using a dispersive qubit-bosonic mode coupling as described in \cite{TW:GKP}. The results in this Section go much beyond \cite{TW:GKP} by using a new phase estimation protocol which assumes no post-selection and which is information-theoretically optimized. 

\subsection{Idea Behind Grid States}
\label{sec:idea}
\begin{figure}[htb]
\includegraphics[width=0.9\hsize]{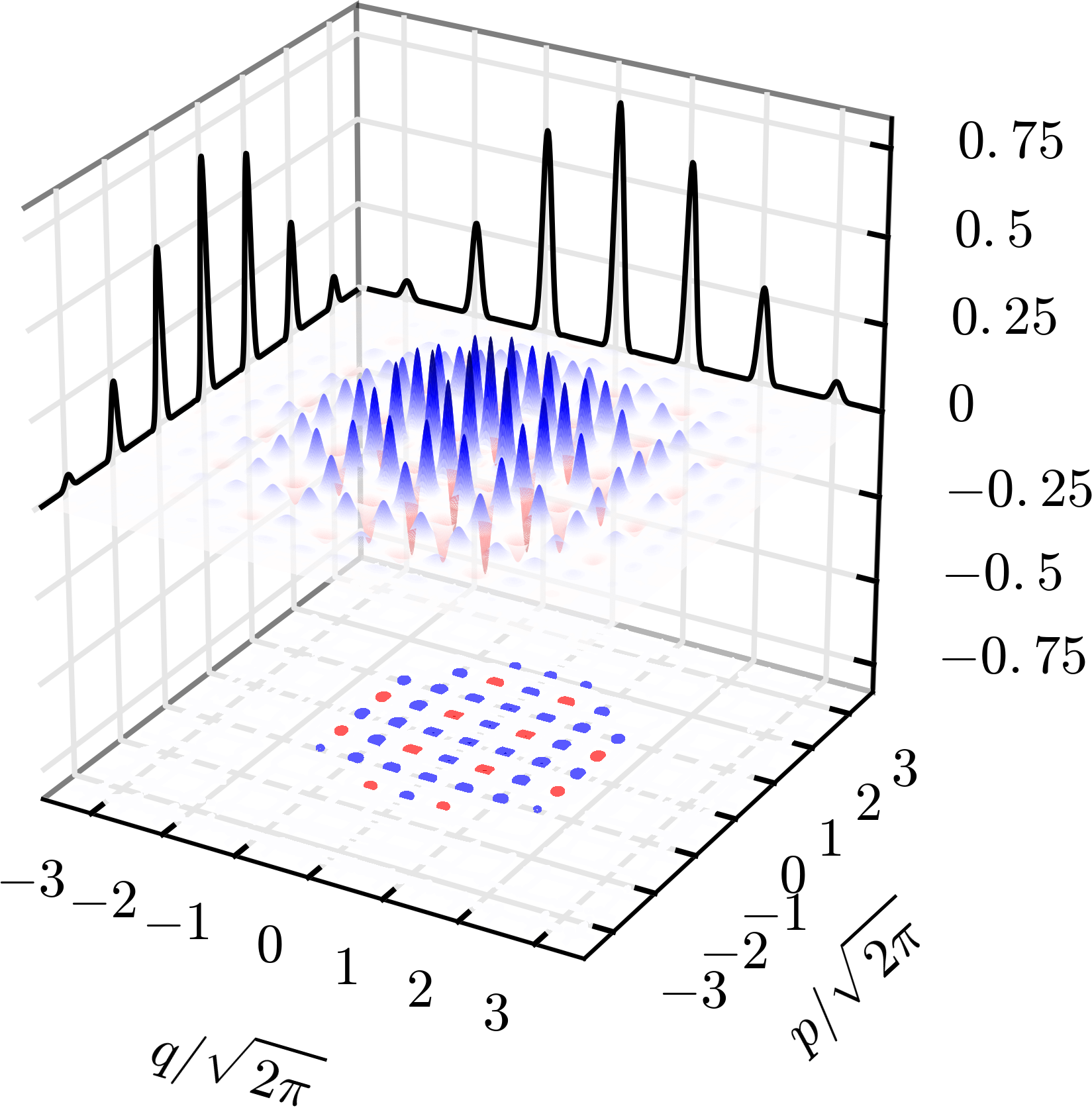}
\includegraphics[width=0.9\hsize]{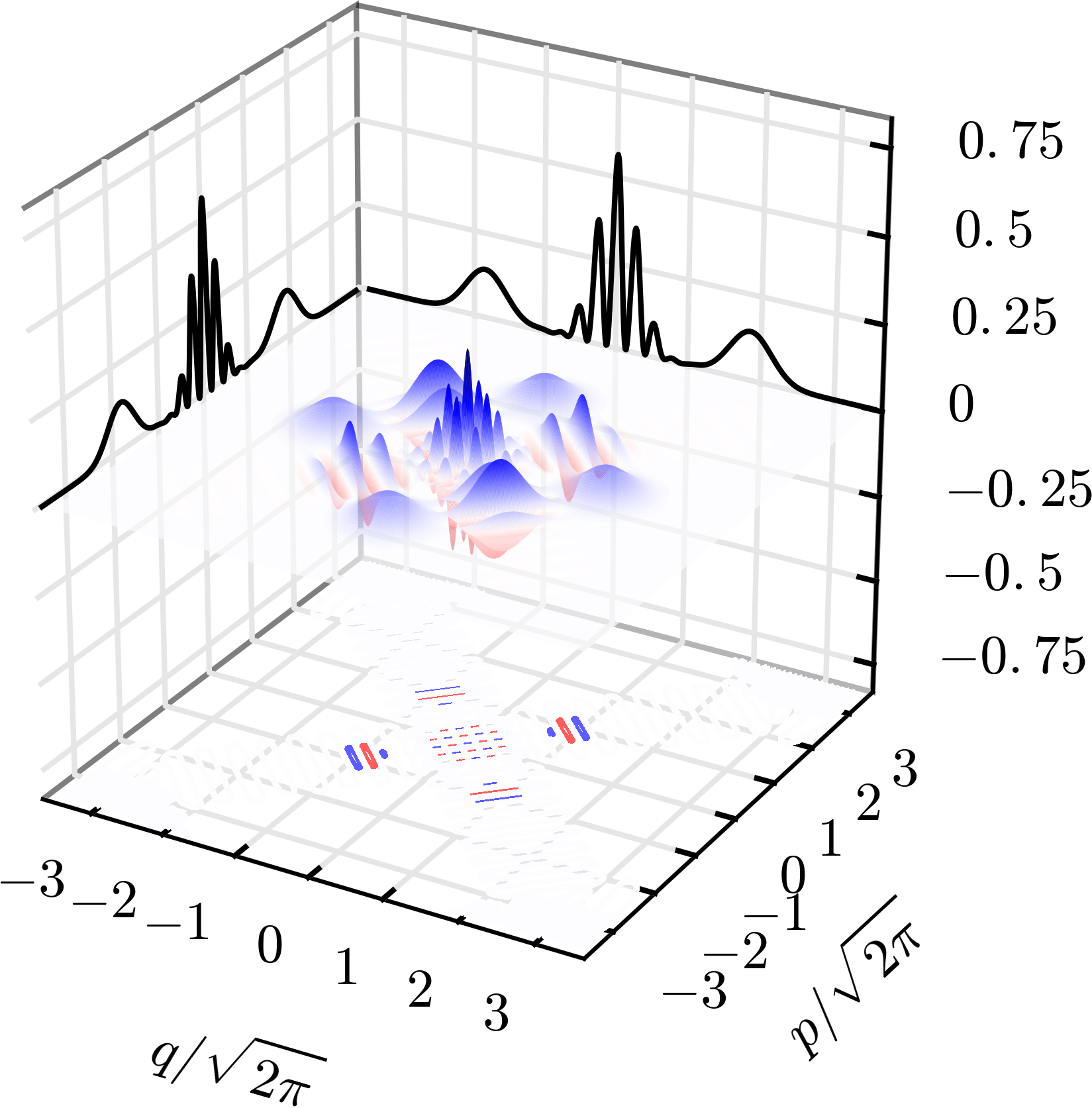}
\caption{Wigner function of a grid state (top) and a quantum compass state (bottom), both with $\overline{n} \approx 12$. The grid state has been generated by the protocol in \cite{TW:GKP} using adaptive phase estimation (corresponding to $l=1$ and an adaptively varying $\varphi$ in Fig.~\ref{fig:SPE}) in $M=8$ rounds, both for $S_p$ and $S_q$ applied to the vacuum state. The distance between the center of the Wigner function peaks is $\sqrt{2\pi}$ in both the $p$ and $q$ direction, thus the unit cell has area $2\pi$. Since the displacement can shift the grid in both positive and negative $p$ and $q$ directions, only displacements which can add at most $\pi/2$ photons to a vacuum state can be distinguished.}
\label{fig:wigner}
\end{figure}

A grid state is an approximate eigenstate of two commuting operators $S_p=e^{i \hat{p}\sqrt{2\pi /(m \omega \hbar)}}$ and $S_q=e^{i \hat{q}\sqrt{2\pi m \omega/\hbar}}$. $S_p$ acts as a shift in position, i.e. $S_p \ket{q}=\ket{q-\sqrt{\frac{2\pi\hbar}{m\omega}}}$ while $S_q$ acts as a shift in momentum, $S_q \ket{p}=\ket{p+ \sqrt{2\pi\hbar m \omega}}$ where $\ket{q}$ and $\ket{p}$ are position and momentum eigenstates. One can verify the commutation of these operators by using the identity $e^A e^B=e^B e^A e^{[A,B]}$ for $A,B$ linear combinations of $\hat{p}$ and $\hat{q}$. From here onwards we redefine $\hat{p}$ and $\hat{q}$ as the dimensionless  `quadrature' operators $\frac{\hat{p}}{\sqrt{m\omega \hbar}}$ and $\hat{q}\sqrt{m\omega/\hbar}$ so that $\hat{p}=\frac{i}{\sqrt{2}}(a^{\dagger}-a)$ and $\hat{q}=\frac{1}{\sqrt{2}}(a+a^{\dagger})$ with canonical commutation relation $[\hat{q},\hat{p}]=i$ (and 
 variances ${\rm Var}(\hat{p})= \frac{1}{2}$ and ${\rm Var}(\hat{q})= \frac{1}{2}$). Hence we use $S_p=e^{i \hat{p} \sqrt{2\pi}}$ and $S_q=e^{i \hat{q} \sqrt{2\pi}}$. A grid state with an average number of photons $\overline{n}$ will not be a perfect eigenstate ($p$ only takes values $k \sqrt{2\pi}$ with $k \in \mathbb{Z}$ and $q=l \sqrt{2 \pi}$ with $l\in \mathbb{Z}$). Rather, its wavefunction in $p$ can be a Gaussian envelope of finite width in which one has a superposition of squeezed peaks at $p=k \sqrt{2\pi}$ and similarly in $q$-space (see an example of the Wigner function in Fig.~\ref{fig:wigner}). 
Now imagine that such grid state undergoes a small displacement $\exp(-i u \hat{p}+i v \hat{q})$ \footnote{Equivalently, the displacement is $D(\beta)=\exp(\beta a^{\dagger}-\beta^* a)$ with $u=\sqrt{2} {\rm Re}(\beta)$ and $v=\sqrt{2} {\rm Im}(\beta)$} (equal to $\exp(-i u\hat{p})\exp(i v\hat{q})$ modulo an overall phase) with unknown parameters $u, v$ in a small interval. The goal is to estimate the value of the parameters $u$ and $v$. In order to determine $u$ and $v$, one measures the eigenvalues of $S_p$ and $S_q$. The shifted grid state $\ket{\psi^{\rm grid}_{u,v}}$ is then an approximate eigenstate of $S_p$ and $S_q$ with eigenvalues $S_p \ket{\psi^{\rm grid}_{u,v}}\approx e^{i \sqrt{2\pi}v} \ket{\psi^{\rm grid}_{u,v}}$ and  $S_q \ket{\psi^{\rm grid}_{u,v}}\approx e^{i\sqrt{2\pi}u} \ket{\psi^{\rm grid}_{u,v}}$. The eigenvalues of $S_p$ and $S_q$ uniquely determine the parameters $u,v$ only when $u,v \in [-\sqrt{\pi/2},\sqrt{\pi/2})$, i.e. the grid should be displaced by less than half the minimum distance between the grid points. A grid state with an increasing number of photons $\overline{n}$ then simply allows for a greater resolution of the eigenvalues.  Grid or comb states were originally proposed by Gottesman, Kitaev and Preskill in \cite{GKP} to represent a qubit in a bosonic mode so that displacement errors can be detected and corrected, but the application for sensing was not considered. The grid state can also be viewed as the phase-space equivalent of the well-known frequency comb in which one has sharp amplitude peaks at equally-spaced frequencies in an overall Gaussian envelope. A form for the finite-photon number grid state is 
\begin{eqnarray}
\ket{\psi^{\rm grid}}=\left(\frac{2}{\pi}\right)^{1/4} \sum_{t=-\infty}^{\infty} e^{-\pi \Delta^2 t^2} \nonumber \\  \int dq \; e^{-(q-\sqrt{2\pi}t)^2/(2\Delta^2)} \ket{q},
\label{eq:grid_def}
\end{eqnarray} 
 which one can read as a squeezed state (in $q$) to which one applies a sum of discrete displacements, weighed with a Gaussian filter. The wavefuntion in $p$ is almost identical, i.e. one can show that
 % DW checked eq, fixed typo
 \begin{equation}
 \ket{\psi^{\rm grid}}=\left(\frac{2}{\pi}\right)^{1/4} \sum_{t=-\infty}^{\infty} \int dp \,e^{-\Delta^2 p^2/2} e^{-(p-\sqrt{2\pi}t)^2/(2\Delta^2)} \ket{p}. \nonumber
 \end{equation}
%DW layout
  The slight asymmetry of the $p$ and $q$ representation vanishes when $\Delta \rightarrow 0$. For this state one can derive that $\overline{n} \approx \frac{1}{4 \Delta^2}$, neglecting $\mathcal{O}(1)$ and $\mathcal{O}(\Delta)$ terms. A grid state is thus squeezed in both quadratures, its sensitivity for detecting displacements can be captured by two effective squeezing parameters which we introduce in Section \ref{sec:squeeze}. 

 % 12.5 $\pm$ 11.5 for grid state. I attached a compass state with 13.3 $\pm$ 3.
\section{Two Parameter Estimation Theory} 
\label{sec:estim}

The task of determining the displacement parameters $u$ and $v$ can be viewed as a problem in quantum estimation theory, see e.g. \cite{helstrom:quantum_CR, yuen_lax, BC:stat_dist, paris:LET, genoni+:displace, book:WM, SBD:multi-parameter}. The relevant question here is: given the pure state $\ket{\psi_{u,v}}=\exp(-i u\hat{p}+i v\hat{q}) \ket{\psi}$ where $\ket{\psi}$ is the sensor state, how well can we estimate $u,v$ using a single quantum measurement? The quantum measurement is given by operation elements $\{E_x\}$ with $\sum_x E_x=I$, leading to outcomes $x$ with probability ${\mathbb P}(x|u,v)={\rm Tr} E_x \ketbra{\psi_{u,v}}{\psi_{u,v}}$. On the basis of an outcome $x$ one chooses an estimator for $u$ as $\tilde{u}(x)$ and for $v$ as $\tilde{v}(x)$. The quantum version of the Cram\'er-Rao bound relates the covariance matrix of the estimators at a point $(u,v)$ to the quantum Fisher information assuming that the estimators $\tilde{u}(x)$ and $\tilde{v}(x)$ are unbiased, meaning that $\sum_x {\mathbb P}(x|u,v)  (\tilde{u}(x)-u)=0,\sum_x {\mathbb P}(x|u,v)  (\tilde{v}(x)-v)=0$ at the point $(u,v)$. In Appendix \ref{sec:2para} we provide a self-contained derivation of this two-parameter quantum Cram\'er-Rao bound for completeness. The bound says that the estimator covariance matrix $\Sigma \geq F^{-1}$ where $F$ is the $2 \times 2$ quantum Fisher matrix whose entries are $F_{ij}=\frac{1}{2}\bra{\psi_{u,v}} L_i L_j+ L_j L_i\ket{\psi_{u,v}}$ with $L_i$ the so-called symmetric  logarithmic derivative operator. For a pure sensor state undergoing a unitary displacement, one can find $L_{i=u}=-2i [\hat{p},\ketbra{\psi_{u,v}}{\psi_{u,v}}]$ and $L_{i=v}=2i [\hat{q},\ketbra{\psi_{u,v}}{\psi_{u,v}}]$, see the tools in \cite{genoni+:displace, boixo+:metro}. As a consequence of the matrix inequality one can obtain the following lower bound on the sum of variances (with ${\rm Var}(\tilde{u})=\sum_x \mathbb{P}(x|u,v) (\tilde{u}(x)-u)^2$ for an unbiased estimator $\tilde{u}$) at a point $(u,v)$:
\begin{equation}
{\rm Var}(\tilde{u})+ {\rm Var}(\tilde{v}) \geq {\rm Tr}(F^{-1}).
\label{eq:lb}
\end{equation}
The Fisher matrix for the displacement problem can be found to be equal to the covariance matrix of the position and momentum observables with respect to the state $\psi_{u,v}$, i.e.$\frac{F_{pp}}{4}={\rm Var}(\hat{p})$, $\frac{F_{qq}}{4}={\rm Var}(\hat{q})$ and
\begin{equation}
\frac{F_{qp}}{4}=\bra{\psi_{u,v}}\hat{p}\ket{\psi_{u,v}}\bra{\psi_{u,v}} \hat{q}\ket{\psi_{u,v}} -\bra{\psi_{u,v}} \frac{1}{2}(\hat{p}\hat{q}+\hat{q}\hat{p}) \ket{\psi_{u,v}}.\nonumber  
\end{equation}

%$\frac{4}{{\rm Tr}(F)}= \frac{1}{{\rm Var}(\hat{p})+{\rm Var}(\hat{q})}$
One can ask for the {\em minimal} value of the lower bound on the r.h.s. of Eq.~(\ref{eq:lb}). For fixed ${\rm Tr}(F)$ we minimize ${\rm Tr}(F^{-1})$ by taking its eigenvalues $\lambda_0=\lambda_1=\frac{{\rm Tr}(F)}{2}$ so that ${\rm Tr}(F^{-1})=\frac{4}{{\rm Tr}(F)}=\frac{1}{{\rm Var}(\hat{p})+{\rm Var}(\hat{q})}$, implying that for any sensor state one has
\begin{equation}
{\rm Var}(\tilde{u})+ {\rm Var}(\tilde{v}) \geq \frac{1}{2\overline{n}+1}.
\label{eq:lowerbound}
\end{equation}
By contrast, for single-mode Gaussian states ((thermal) squeezed or coherent states), Ref.~\cite{genoni+:displace} has shown that ${\rm Var}(\tilde{u})+ {\rm Var}(\tilde{v}) \geq 2$. In addition, for a pure two-mode squeezed state, in the limit of large squeezing, Ref.~\cite{genoni+:displace} obtains the same lower bound as in Eq.~(\ref{eq:lowerbound}), which the superdense coding protocol effectively achieves.
% BMT: if one takes the lower bound 1/Var(p) + 1/Var(q), infinite squeezing gives one 2. 
% two-mode squeezing, they get lower bound of 2 exp(-2r) and sinh^2(r)= overline(n)
The bound shows that it {\em may} be possible to suppress the variances of both displacement parameters simultaneously for increasing photon number. One can observe that for the quantum compass state, the grid state and a photon number state $\ket{n}$ at the symmetry point $u=v=0$, ${\rm Tr}(F^{-1})=4/{\rm Tr} F=\frac{1}{2\overline{n}+1}$ since $F_{qp}=0$: at this point $\bra{\psi_{0,0}} \hat{p} \hat{q} \ket{\psi_{0,0}}=-\bra{\psi_{0,0}} \hat{q} \hat{p} \ket{\psi_{0,0}}$ due to the symmetry of these states in phase space (see the invariance under the phase space rotation $\hat{p} \rightarrow \hat{q}, \hat{q}\rightarrow -\hat{p}$ in Fig. \ref{fig:wigner}): all these states have in common that they maximize the uncertainty in $p$ and $q$ and do so symmetrically. However, it is not known whether in the two-parameter case the bound in Eq.~(\ref{eq:lowerbound}) is tight (see discussion in the Appendix). Between achieving the minimal lower bound in Eq.~(\ref{eq:lowerbound}) and being stuck with variances which cannot be driven down by increasing photon numbers, there is substantial room for interesting behavior which we will now discuss.

\section{Phase Estimation Protocol for Grid States}
\label{sec:estimphase}
\begin{figure}[htb]
\centering
\parbox{3cm}{
\Qcircuit @C=1em @R=1em {
\lstick{\mbox{oscillator}} & \qw & \qw & \gate{U^l} & \qw & \qw & \qw & \qw \\
\lstick{\mbox{qubit} \ket{0}} & \gate{H} & \qw & \ctrl{-1} &  \gate{R_z(\varphi)} & \gate{H} & \meter
}}
\caption{Phase estimation for a unitary operator $U$, for example $U=S_p=e^{i \sqrt{2\pi}p}$ or $S_q$ consists of applying the `Ramsey' circuit of this Figure for $M$ rounds, with possibly varying $l$ and phases $\varphi$ repeatedly on one oscillator input state. If the input state is an eigenstate of $U$ with eigenvalue $e^{i \theta}$, then the probability for outcome 0 for the qubit measurement is given by ${\mathbb P}(0|\theta)=\frac{1}{2}(1+\cos(l\theta +\varphi))$. In textbook phase estimation with $M$ ancilla qubits, one takes one takes $l=2^k$ with $k=M-1, \ldots, 0$ starting at $k=M-1$. The rotation $R_z(\varphi)$ depends on the outcomes of previous qubit measurements (feedback) so the circuits effectively implement the textbook quantum Fourier transform in sequential manner. In the phase estimation used in Section \ref{sec:exp} we use $l=1$ and the feedback phases $\varphi$ are chosen to maximize the information gain about the phase.}
\label{fig:SPE}
\end{figure}
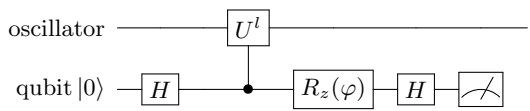

The (approximate) measurement of the eigenvalue $e^{i\theta}$ of a unitary operator $U$, and the simultaneous projection of the input state onto the corresponding eigenstate $\ket{\psi_{\theta}}$ with $U\ket{\psi_{\theta}}=e^{i\theta}\ket{\psi_{\theta}}$, is called phase estimation for $U$. One can thus use phase estimation for the operators $S_p$ and $S_q$ applied to a vacuum input state to prepare an approximate eigenstate of these operators, a grid state $\ket{\psi^{\rm grid}}$ as was shown in detail in \cite{TW:GKP}. After preparation, a displacement with parameters $u$ and $v$ happens and one executes the same phase estimation for $S_p$ and $S_q$ to measure the change in eigenvalues.  A phase estimation protocol can be executed by repeatedly (and sequentially) coupling the oscillator state to a qubit as depicted  in Fig.~\ref{fig:SPE}.
 We have shown in \cite{TW:GKP} that for a $M$-round simple repeated phase estimation protocol with $l=1$ and fixed phases $\varphi=0$ for $M/2$ rounds and $\varphi=\pi/2$ for $M/2$ rounds in Fig.~\ref{fig:SPE}, as well as for textbook phase estimation, one gets a phase variance ${\rm Var}(\theta) \sim \frac{1}{\overline{n}}$.  This scaling is also consistent with the abstract description of a grid state in terms of the squeezing parameter $\Delta \ll 1$ in Eq.~(\ref{eq:grid_def}).  Each wavefunction peak in $q$ (resp. $p$) under the Gaussian envelope with standard deviation $\frac{1}{\Delta}$ has ${\rm Var}(p) \sim \Delta^2$ (resp. ${\rm Var}(q) \sim \Delta^2$) and given that $\overline{n} \approx \frac{1}{4 \Delta^2}$ one expects that ${\rm Var}(\theta)\sim \Delta^2\sim\frac{1}{\overline{n}}$. 

These scaling arguments suggest that it may be possible to achieve $\frac{1}{\overline{n}}$ scaling for the sum of variances in Eq.~(\ref{eq:lowerbound}). In Appendix \ref{sec:phase} we present a fully rigorous analysis which shows that one can at least achieve $\frac{1}{\sqrt{n}}$ scaling if one uses textbook phase estimation for the preparation and (displacement) measurement of the grid state. The difference between this measurement analysis and the scaling arguments given above is that the latter is a fully rigorous analysis of the variances for the estimates $\tilde{u}$ and $\tilde{v}$, while the scaling arguments above pertains to the variances of $p$ and $q$ in the wave-function of the state. Even though the choice of texbook phase estimation preparation may not be the most efficient in practice, it is the only explicit preparation protocol for which this kind of analysis seems analytically feasible. The noisy protocols in Section \ref{sec:exp} use a much more photon-efficient and suitable form of phase estimation. 
 
More precisely, we show in the Appendix that one can obtain estimators $\tilde{u}$ and $\tilde{v}$ for which it holds that the sum of the mean square deviations ${\rm MSD}(\tilde{u})+{\rm MSD}(\tilde{v}) = O(\frac{1}{\sqrt{\overline{n}}})$ with ${\rm MSD}(\tilde{v})=\sum_{\tilde{v}}\mathbb{P}(\tilde{v} |v) (\tilde{v}-v)^2$. As the estimator is slightly biased, one uses the MSD instead of the variance.
 For textbook phase estimation one can get an exact expression for the probability $\mathbb{P}(\tilde{v} |v)$ (and identically $\mathbb{P}(\tilde{u} |u)$) (see Appendix), namely
\begin{equation}
{\mathbb P}(\tilde{v}|v)=\frac{1}{2^{2M}} \frac{ \sin^2(2^M \sqrt{\pi/2}(\tilde{v}-v))}{\sin^2(\sqrt{\pi/2}(\tilde{v}-v))},
\label{eq:expP}
\end{equation}
where $\frac{\tilde{v}}{\sqrt{2\pi}}=-\frac{1}{2}+\frac{y}{2^M}$ with integers $y\in \{0,1\}^M$ where $M$ is the number of rounds, see Fig.~\ref{fig:SPE}. The number of photons in a grid state prepared using $M$-round phase estimation for $S_p$ and $M$-round textbook phase estimation for $S_q$ scales $\overline{n} \sim 2^{2M}$ as each round with controlled-displacement $S_p^{2^k}$ (or $S_q^{2^k}$) adds $O(2^{2k})$ photons.

Armed with Eq.~(\ref{eq:expP}) we can prove that ${\rm MSD}(\tilde{v}) =O(\frac{1}{2^M})$ for values of $v$ inside an interval $I=[-\sqrt{\frac{\pi}{2}}+2\alpha \sqrt{\frac{\pi}{2}},\sqrt{\frac{\pi}{2}}-2\alpha \sqrt{\frac{\pi}{2}}]$ with a constant $0<\alpha< \frac{1}{2}$, hence scaling as 
$1/\sqrt{\overline{n}}$. The interval $I$ is necessary so that large errors in the parameter occuring at the boundary of the periodic interval are avoided. The bias of the estimators $\tilde{v}, \tilde{u}$ makes the difference between the MSD and variance in fact negliglible, of $O(2^{-2M})=O(\overline{n}^{-1})$. Details of this derivation are given in the Appendix.

Unlike for the grid state we do not see an obvious way to obtain such $\frac{1}{\sqrt{\overline{n}}}$ or even $\frac{1}{\overline{n}}$ scaling for the sum of variances for a quantum compass state nor for a photon number state sensor  \footnote{For a displaced photon number state sensor $D(\beta) \ket{n}$, a measurement in the photon number basis would not allow one to resolve small displacements $|\beta|^2 \ll 1$. A measurement in the overcomplete basis $\ket{\psi_{\tilde{\beta}}}=D(\tilde{\beta})\ket{n}$ would output the estimate $\tilde{\beta}$ with probability $\mathbb{P}(\tilde{\beta}|\beta)=\frac{1}{\pi} |\bra{n} D(\beta-\tilde{\beta}) \ket{n}|^2=\frac{1}{\pi}
e^{-|\beta-\tilde{\beta}|^2}(L_n(|\beta-\tilde{\beta}|^2))^2$ with Laguerre polynomial $L_n(x)$, whose support for $x > 0$ increases for larger $n$.}.

% u+iv = sqrt(2) beta with D(beta)
% for gamma=u+iv ${\rm Var}(\gamma)=\langle (\gamma-\langle \gamma \rangle)(\gamma^*-\langle \gamma^* \rangle)\rangle$= Var(u)+Var(v)
% so Var(beta)=1/2 (Var(u)+Var(v)) 
%  inequalities are for not assuming any prior distribution
% QFI analysis
% BMT does not take ensemble of displacements into account, no a priori probability distribution,as it captures sensitivity of p(data |u,v) and  u_est=F(data), v_est=G(data) to changes in u and v.
%\footnote{There are versions of this bound which do take into account the probability distribution over $u$ and $v$, see \cite{paris:LET}.}
% other evidence

%DW fixed caption: top/bottom
\begin{figure}[htb]
\includegraphics[width=1\hsize]{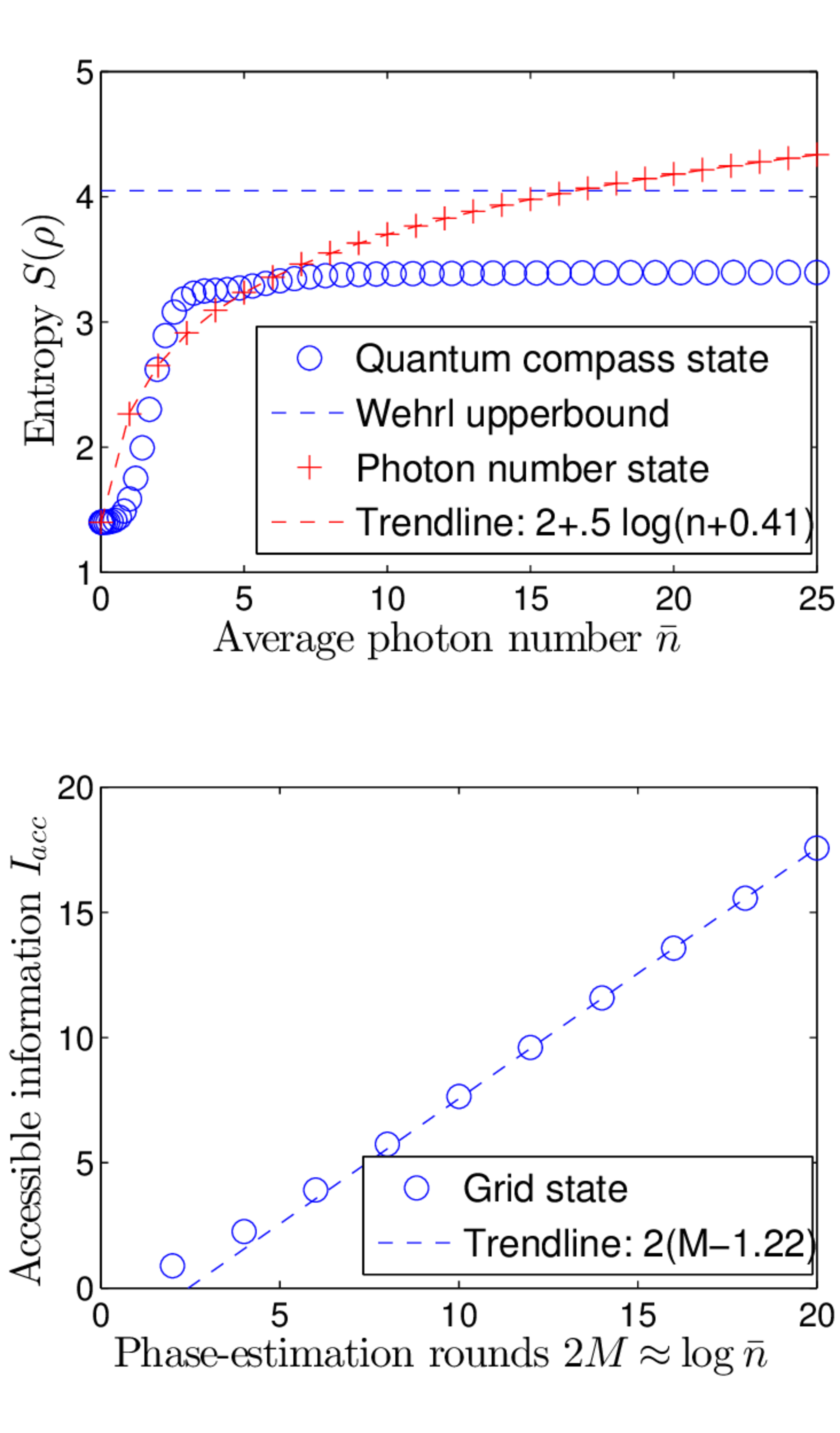}
\caption{Top: The growth of the Von Neumann entropy $S(\rho)$ with the number of photons $\overline{n}$ in a quantum compass state which is displaced by a small random amount and a constant, photon number independent, Wehrl entropy upperbound for $S(\rho)$ derived in \ref{sec:wehrl}. Bottom: the growth of the accessible information in the grid state as a function of $2M \approx \log (\overline{n})$.}
\label{fig:entropy_bound}
\end{figure}

%Eq.~(\ref{eq:condprob}) shows that $\mathbb{P}(\tilde{v}|v)$ only depends on $|\tilde{v}-v|$ and thus $\sum_{\tilde{v}} (\tilde{v}-v) \;\mathbb{P}(\tilde{v} |v)=0$, implying that the estimator is unbiased for any $v$.

% BMT one can evaluate Fisher information for grid states, showing how sensitive the state is
% similar for quantum compass state?

\section{Information-Based Comparison with Quantum Compass State} 
\label{sec:qcompass}

A complementary perspective is offered by analyzing the information about the displacement in the sensor state assuming some probability distribution over the displacement parameters. The reason to do this is simple: it is not known whether the r.h.s. in Eq.~(\ref{eq:lb}) is achievable by a measurement, hence the Cram\'er-Rao lower bound does not allow one to show how poorly other sensor states perform. However if we assume a probability distribution over possible displacements then one can consider the accessible information in a state about the displacement and one can upperbound the accessible information by the von Neumann entropy (and use upper bounds on the von Neumann entropy).

We choose the displacement parameter $\beta$ to be uniformly distributed in a {\em small} constant interval, namely ${\rm Re}(\beta),{\rm Im}(\beta) \in [-\sqrt{\pi}/2,\sqrt{\pi}/2)$ (directly corresponding to $u$ and $v \in [-\sqrt{\pi/2},\sqrt{\pi/2})$). The mutual information about the parameters $u,v$ when a grid state is prepared and measured using phase estimation equals $I_{\rm acc}=2M +  \frac{2}{\sqrt{2\pi}} \int_{-\sqrt{\pi/2}}^{\sqrt{\pi/2}} dv \sum_{\tilde{v}} {\mathbb P}(\tilde{v}|v) \log {\mathbb P}(\tilde{v}|v)$ with $\mathbb{P}(\tilde{v}|v)$ in Eq.~(\ref{eq:expP}) (see details in Appendix \ref{sec:phase}). In Fig.~\ref{fig:entropy_bound} we plot $I_{\rm acc}$ as a function of $2M\approx \log(\overline{n})$ where $M$ is the number of qubits/rounds used in phase estimation. The plot can be used to show that $I_{\rm acc}$ grows as $2M-c$ with constant $c$, hence as $\log(\overline{n})$, showing that one can resolve the displacement parameters with higher resolution for increasing photon numbers.

%DW Question: The constant bound is mentioned twice, is that intentional?(Sentences 1 and 3 are very similar)
For a quantum compass state we can upper bound the accessible information $I_{\rm acc}$ in the state about the displacement by {\em a constant}, independent of photon number, see Fig.~\ref{fig:entropy_bound}. We use the Holevo bound $I_{\rm acc}\leq S(\rho)$ where $\rho=\int_S d\beta D(\beta)\ketbra{\psi^{\rm comp}}{\psi^{\rm comp}} D(-\beta)$ where $S$ is the integration region for $\beta$. In the next section we present a constant {\em analytic} upper bound on the entropy $S(\rho)$ which holds in the limit of large photon number $\overline{n}$. We can also compare the information in the grid state to displacement information when we choose a photon number state $\ket{n}$ as a sensor state. Again we upper bound the accessible information about the displacement by the von Neumann entropy of $\rho_n= \int_S d\beta D(\beta)\ketbra{n}{n} D(-\beta)$, see Fig.~\ref{fig:entropy_bound}. For the displaced photon number state it is unclear whether there exists a measurement through which one gains this amount of information $S(\rho_n)$.

%BMT analysis does not depend on distribution over v, Gaussian also ok.
\subsection{Upper Bound on Von Neumann Entropy for Quantum Compass State}
\label{sec:wehrl}
 In order to upper bound the von Neumann entropy $S(\rho)=-{\rm Tr} \rho \log \rho=-\frac{1}{\pi} \int d\gamma \bra{\gamma} \rho \log \rho \ket{\gamma}$ we use the Wehrl entropy $S_W(\rho)=-\frac{1}{\pi}\int d\gamma f_{\rho}(\gamma) \log f_{\rho}(\gamma)\geq 0$ where $f_{\rho}(\gamma)=\bra{\gamma} \rho \ket{\gamma}$.  One has $S_W(\rho)=-\frac{1}{\pi} \int d\gamma \bra{\gamma} \rho \ket{\gamma} \log \bra{\gamma} \rho \ket{\gamma}\geq -\frac{1}{\pi} \int d\gamma \bra{\gamma} \rho \log \rho \ket{\gamma}=-{\rm Tr}\rho \log \rho=S(\rho)$ \cite{wehrl:1979}.
In the next few simplications we use that $\overline{n} \gg \frac{\pi}{2} \geq |\beta|^2$ (where $\beta$ is the displacement parameter) and we omit terms which will vanish when $\overline{n}$ grows. In this limit, $\bra{\gamma} \rho\ket{\gamma}$ only depends on the diagonal entries of $\rho$, i.e. 
we approximate $S_W(\rho) \approx S_W(\rho_{\rm diag})$ where  
%DW Layout
\begin{eqnarray}
\lefteqn{\rho_{\rm diag}=\frac{1}{4}\int_S d\beta D(\beta)} \nonumber \\
& & (\ketbra{\alpha}{\alpha}+\ketbra{-\alpha}{-\alpha}+\ketbra{i\alpha}{i\alpha}+\ketbra{-i \alpha}{-i \alpha})D(-\beta). \nonumber
\end{eqnarray}
Here $\int_S d\beta=\frac{1}{\pi} \int_{-\sqrt{\pi}/2}^{\sqrt{\pi}/2} \int_{-\sqrt{\pi}/2}^{\sqrt{\pi}/2} d{\rm Re}(\beta)\ d{\rm Im}(\beta)$. 
The integral $\int d\gamma$ in $S_W(\rho_{\rm diag})$ can be broken up in four regions of phase space, i.e. North (N), South (S), East (E), West (W) such that each region contains only the point $\alpha$ (E), $-\alpha$ (W),$ i\alpha$ (N), $-i \alpha$ (S). 

Let $\rho_{\alpha}=\int_S d\beta D(\beta) \ket{\alpha}\bra{\alpha} D(-\beta)$. In the large photon limit we have 
that for $\gamma \in {\rm N},{\rm S}, {\rm W}\colon|\bra{\gamma} \rho_{\alpha} \ket{\gamma}| \approx 0$ as the total support of $\rho_{\alpha}$ will be well contained in the region E. A similar statement holds for $\rho_{\pm i\alpha},\rho_{-\alpha}$. Hence one may approximate the entropy $S_W(\rho_{\rm diag})$ by 
the sum of four separate and identical contributions
\begin{eqnarray}
S_W(\rho_{\rm diag}) & \leq & -\frac{1}{\pi}\int d\gamma \bra{\gamma} \rho_{\alpha} \ket{\gamma} \log \left(\frac{\bra{\gamma} \rho_{\alpha}\ket{\gamma}}{4}\right) \nonumber \\ & = & 2+S_W(\rho_{\alpha}) \nonumber.
\end{eqnarray}
Since we integrate over all of phase space to determine $S_W(\rho_{\alpha})$, it is in fact irrelevant that the state is centered around some $\alpha$, i.e. we switch variables $\gamma \rightarrow \gamma+\alpha$ and use $\int d (\gamma+\alpha)=\int d\gamma$.
Thus the following bound holds for large $\overline{n} \gg \pi/2$:
\begin{eqnarray}
S(\rho)\lesssim -\frac{1}{\pi}\int d\gamma f(\gamma) \log f(\gamma)+2, \nonumber \\
f(\gamma)=\int_S d\beta \exp(-|\beta-\gamma|^2).
\nonumber
\end{eqnarray}
We plot this constant upper bound in Fig.~\ref{fig:entropy_bound}. 

\section{Numerical Analysis of Noise During Preparation and Measurement}
\label{sec:exp}

A high-coherence superconducting transmon qubit dispersively coupled to a high-Q microwave cavity is a good candidate for an experimental realization of a grid state as was argued in \cite{TW:GKP}. A single round of phase estimation has (almost) identical experimental components as a photon parity measurement performed in \cite{ofek+:QEC,sun+:parity} (where such measurement rounds are performed repeatedly). Protocols with $M=4-8$ of rounds of non-adaptive and adaptive phase estimation which save on photon use as compared to textbook phase estimation are described in \cite{TW:GKP}. Note that the number of photons in the sensor state made in $M$ rounds is half that of an $M$-round encoded state in \cite{TW:GKP} as the two displacement operators $S_p$ and $S_q$ for the sensor {\em state} are weaker in strength than the displacement checks which define a code {\em space}. An alternative platform could be the creation of a grid state in a microwave cavity by sequentially passing Rydbergh atoms through the cavity \cite{book:haroche} implementing the 4-8 rounds of Ramsey phase estimation in Fig.~\ref{fig:SPE}. The advantage of this set-up is that the cavity-atom interaction is only `on' while the atom is in transit through the cavity, but the feasibility of this scheme has not yet been analyzed. For superconducting transmon qubits coupled to microwave cavities, the preparation protocols that we consider may not even be needed as it is possible to create any state of the cavity by numerically optimizing microwave pulses on cavity and qubit system \cite{heeres+:control}. 

In this section we analyze a phase estimation preparation protocol which is based on a dispersive qubit-cavity interaction of the form $Z a^{\dagger} a$. We first introduce our performance measure which captures how well one prepares a grid state. We then discuss our choice of information-gain-optimized phase estimation and present the Hamiltonian and simulated noise models.

\subsection{Quality Measure: Effective Squeezing Parameters}
\label{sec:squeeze}

Given a probability distribution ${\mathbb P}(\theta)$ where $\theta\in [-\pi,\pi)$, the Holevo phase variance $\sigma_H$ is defined as $\sigma_H=\sqrt{|\braket{e^{i\theta}}|^{-2} - 1}$ with $\langle e^{i\theta} \rangle=\int_{-\pi}^{\pi} d\theta \;{\mathbb P}(\theta) e^{i \theta}$. 
For a state $\rho$ and an operator $S_p$ one has ${\rm Tr} S_p \rho= \langle e^{i\theta} \rangle$ where $e^{i\theta}$ is the possible eigenvalue of $S_p$ and ${\mathbb P}(\theta)=\bra{\psi_{\theta}} \rho \ket{\psi_{\theta}}$. Thus the Holevo phase variance could in principle measure how well the state $\rho$ is an eigenstate of $S_p$, see \cite{TW:GKP}. Here we will use a slightly different measure: given a state $\rho$ and the operator $S_p$ (similarly $S_q$) we will consider {\em the effective squeezing parameter} defined as
\begin{equation}
\Delta_p(\rho)\equiv  \sqrt{\frac{1}{\pi} \ln\left(\frac{1}{|{\rm Tr}S_p \rho |^2}\right)}.
\end{equation}
For a squeezed vacuum state $\ket{\rm sq. vac.}$ in $p$ with squeezing parameter $\Delta <1$ such that ${\rm Var}(p)=\frac{1}{2}\Delta^2$, one has $\bra{\rm sq. vac.} S_p\ket{\rm sq. vac.} = e^{-\frac{\pi}{2} \Delta^2}$ from which it follows that $\Delta({\rm sq. vac})=\Delta$.
Each state $\rho$ can thus be characterized by two effective squeezing parameters $\Delta_p(\rho)$ and $\Delta_q(\rho)$. For the grid state in Eq.~(\ref{eq:grid_def}) one has $\Delta_p \approx \Delta_q=\Delta$. For a coherent state one has $\Delta_p=\Delta_q=1$. 
% BMT Daniel please verify true for both quadratures? For the grid state i wrote Delta_p \approx \Delta, only is true strictly when Delta becomes very small i think, has to do with asymmetry in the state

The motivation for these parameters is as follows. We expect that for an approximately prepared grid state $\rho$ the distribution ${\mathbb P}(\theta)$ is close to that of a {\em wrapped} Gaussian distribution ${\mathbb P}_G(\theta)$ with mean $\mu$ and standard deviation $\sigma$:
\begin{align*}
-\pi \leq \theta < \pi,\; {\mathbb P}_G(\theta) &= \sum_{n=-\infty}^\infty \frac{1}{\sqrt{2\pi\sigma^2}} e^{\frac{-(\theta-\mu + 2\pi n)^2}{2\sigma^2}}.
\end{align*}
One can easily prove that for such a wrapped Gaussian distribution ${\mathbb P}_G(\theta)$, one has
\begin{align*}
\braket{e^{i\theta}} &= \int_0^{2\pi}\mathrm{d}\theta\ e^{i\theta} \sum_{n=-\infty}^\infty \frac{1}{\sqrt{2\pi\sigma^2}} e^{-\frac{(\theta-\mu + 2\pi n)^2}{2\sigma^2}},\\
&= \int_{-\infty}^{\infty}\mathrm{d}\theta\ e^{i\theta} \frac{1}{\sqrt{2\pi\sigma^2}} e^{\frac{-(\theta-\mu)^2}{2\sigma^2}},\\
&= e^{i\mu -\frac{\sigma^2}{2}}.
\end{align*}
or
\begin{align}
&\mu = \arg \braket{e^{i\theta}}, && \sigma = \sqrt{\ln\left(\frac{1}{|\braket{e^{i\theta}}|^2}\right)}.
\label{eq:estimate_std}
\end{align}
This implies that the squeezing parameters of this distribution directly relate  to the standard deviation of the Gaussian distribution and thus $\arg {\rm Tr} S_p \rho$ gives a good estimate for the phase. When $|{\rm Tr}S_p\rho|$ is close to 1, one can use $\ln(1+x)= x+\mathcal{O}(x^2)$ to show that $\Delta_p(\rho)=\frac{1}{\sqrt{\pi}} \sqrt{|{\rm Tr}S_p \rho|^{-2} -1}$, thus relating to the Holevo phase variance.

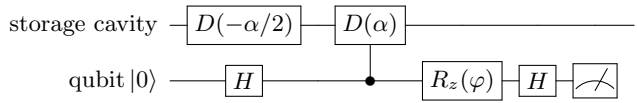
\begin{figure*}[htb]
\begin{minipage}{\textwidth}
	\Qcircuit @C=.7em @R=.7em {
		\leftlabel{\mbox{storage cavity}} & \gate{D(-\alpha/2)} & \qw & \gate{D(\alpha)} & \qw & \qw & 			\qw & \qw \\
		\leftlabel{\mbox{qubit} \ket{0}} & \gate{H} & \qw & \ctrl{-1} &  \gate{R_z(\varphi)} & 					\gate{H} & \meter
	}
\end{minipage}
	\caption{Phase estimation circuit for preparing a grid state. One has $\alpha=\sqrt{\pi}$ for a measurement of $S_p$ and $\alpha=i\sqrt{\pi}$ for $S_q$.}
	\label{fig:PE_GKP}
\end{figure*}
% BMT Daniel alpha in D(alpha) were for GKP code not displacement sensor, please verify
% DW fixed

\begin{figure*}[htb]
\begin{minipage}{\textwidth}
	\Qcircuit @C=.7em @R=.7em {
		\leftlabel{\text{cavity}} & \multigate{2}{R(-Z\pi/2)} & \gate{D(-i\alpha/2)} & \multigate{2}{R(-Z\pi/2)} &
			\qw & \qw &
			\push{\hspace{0.7em}} &  & &
			\multigate{2}{D(-Z\alpha/2)} &	\qw &
			\\	
		&&&&
		&&
		\push{\hspace{0.7em}} & = &&\\	
		\leftlabel{\text{qubit}}  & \ghost{R(-Z\pi/2)} & \gate{X}  &\ghost{R(-Z\pi/2)}  &
			\gate{X} & \qw &
			\push{\hspace{0.7em}} &  & &
			\ghost{D(-Z\alpha/2)} & \qw &
	}
\end{minipage}
	\caption{Equivalent circuit for $D(-Z\alpha/2)$, here $R(-Z\pi/2)=\exp(iZ a^{\dagger} a\pi/2)$.}
	\label{fig:Controlled_Disp}
\end{figure*}
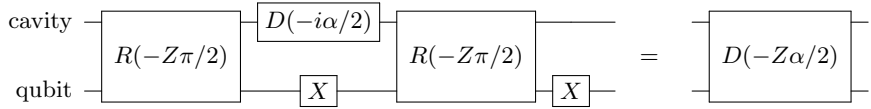

\subsection{Choice of Phase Estimation}
\label{sec:choice}
For phase estimation, we will only use the circuit in Fig.~\ref{fig:SPE} with $l=1$.  Many variants of phase estimation could be considered, e.g. going beyond Fig.~\ref{fig:SPE} by entangling qubits between rounds or performing joint measurements on qubits, but simplicity is what we opt for here. Our only choice for optimization is thus the choice of feedback phase $\varphi$ which can depend on outcomes of previous qubit measurements. A strategy which is then performing better than the one that we originally chose in \cite{TW:GKP}, in terms of the achieved $\Delta(\rho)$, is the following maximal-information gain strategy. 

Given a sequence of qubit measurement outcomes $x_1,\ldots, x_{M-1}$, assume that one somehow determines an estimate $\tilde{\theta}_{M-1}$ for the phase $\theta$. Given this estimate $\tilde{\theta}_{M-1}$ we will then choose the next circuit with a feedback phase $\varphi$ such that the probability for qubit outcome 0 and qubit outcome 1 are equally likely. Hence, given what we believe we know, we choose our next `Ramsey experiment' such that we gain maximal information of 1 bit. If the input to the phase estimation circuits is the eigenstate $\ket{\psi_{\theta}}$, then the 
probability for outcome $x$ equals ${\mathbb P}(x)=\frac{1}{2}(1+(-1)^x\cos(\theta +\varphi))$ so that the maximal-information gain condition simply reads
\begin{align*}
\varphi_M = \tilde{\theta}_{M-1} + \frac{\pi}{2}.
\end{align*}
%DW 'using any integer l' instead of '$l\neq 1$'
Note that this strategy can be employed in any phase estimation, using any integer $l$, which proceeds by several rounds of circuits of the form in Fig.~\ref{fig:SPE}: one can always get a current estimate of the phase given the data and make sure that the next measurement with some number of outcomes gives equal probability to all these outcomes. Textbook phase estimation based on the quantum Fourier transform follows this strategy as well: the lowest significant bit of the phase is estimated with a binary measurement first, then the feedback phase is adapted such that the next measurement comes out 0 or 1 with probability $1/2$, only depending on the next significant bit etc.

In order to estimate the current phase $\tilde{\theta}_{M-1}$ based on the measurement outcomes $x_1,\ldots,x_{M-1}\equiv {\bf x}_{M-1}$ we observe the following. Given ${\bf x}_{M-1}$ and an input state to the protocol one can calculate the output state $\rho({\bf x}_{M-1})$. The description of $\rho({\bf x}_{M-1})$
can include the noise in the protocol, e.g. measurement errors, in order to best describe the state that the protocol produces. Motivated by Eq.~(\ref{eq:estimate_std}) it is then natural to take 
\begin{equation}
\tilde{\theta}_{M-1}=\arg {\rm Tr} S_p \rho({\bf x}_{M-1}).
\end{equation}
%DW made more precise
In the noiseless case, this choice is merely equivalent to 
\begin{equation}
\tilde{\theta}_{M-1}=\arg \int d\theta {\mathbb P}(\theta| {\bf x}_{M-1}) e^{i \theta},
\end{equation}
with ${\mathbb P}(\theta|{\bf x}_{M-1})=\frac{{\mathbb P}({\bf x}_{M-1}| \theta) {\mathbb P}(\theta)}{{\mathbb P}({\bf x}_{M-1})}$. In our simulations we assume ${\mathbb P}(\theta)$ is uniform, which is warranted for our input state. The fact that ${\mathbb P}({\bf x}_{M-1})$ is independent of $\theta$ then allows for the simple expression 
$\tilde{\theta}_{M-1}=\arg \int d\theta\; {\mathbb P}({\bf x}_{M-1}| \theta) e^{i \theta}$, which is also what we have used previously \cite{TW:GKP}. In our simulations we have found that using the noisy $\rho({\bf x}_{M-1})$ to estimate the current phase always gives better or equal results (depending on the setting, $\Delta$ can change up to $0.01$), but in our plots we always use the state produced by the noiseless protocol to estimate the current phase.

%In our simulations we thus do not estimate $\tilde{\theta}_{M-1}}$ via calculating the noisy $\rho({\bf x}_{M-1})$.

\subsection{Hamiltonian and Noise Model}

%DW language
A single round in Fig.~\ref{fig:SPE} can be implemented using the circuit in Fig.~\ref{fig:PE_GKP} which uses controlled-displacements (and cavity pre-displacement). We will assume that the qubit that is measured is again used in the next round. If we imagine using the qubit-cavity interaction $\chi Z a^{\dagger} a$, one implements  Fig.~\ref{fig:PE_GKP} using Fig.~\ref{fig:Controlled_Disp}. 

In order to include nonlinearities, photon loss, qubit decay and measurement errors in the simulation we use a simplified model for when these unwanted processes act. Namely we let them operate during the operations which take the longest to enact in practice.

We will assume that all gates which act either on the qubit or on the cavity mode are very fast compared to all other timescales: we approximate them as taking place instantaneously. The cavity-qubit rotation $R(-Z\pi/2)$ gates take a time $\pi/2\chi \approx 104\ \mathrm{ns}$ each when $\chi/2\pi = 2.4\ \mathrm{MHz}$.  In the simulation it is assumed that cavity and qubit can be decoupled, i.e. $\chi$ can be turned `off' when this interaction should not take place. The measurement process is modeled as an instantaneous (possibly faulty) projection followed by a finite readout idling time of $150\ \mathrm{ns}$. The outcome of the projection is used to reset the qubit to $\ket{0}$ for the next round (assuming a noiseless instantaneous Pauli $X$ gate). However, during the measurement idling time the qubit can still decay from $\ket{1}$ to $\ket{0}$ which would imply that we start the next round with the qubit in a wrong state. When we show data for qubit amplitude damping we thus include this error process. The time-scale of a single phase estimation round is thus determined by $\chi$ and the measurement time, both of which are taken to be reasonable values of superconducting transmon qubit experiments \cite{ofek+:QEC}. 

% BMT w_q Z/2 and w_r a^dagger a not included in simulation
\subsubsection{Nonlinearity}
We have previously identified nonlinearities as a possible cause for bad errors for a GKP state \cite{TW:GKP}: we thus simulate the effect of two different nonlinearities. One is the cavity anharmonicity, also called Kerr interaction,
\begin{align*}
H_{Kc} = -\frac{K_{c}}{2}a^{\dag 2} a^2.
\end{align*}
The other is the non-linear dispersive shift
\begin{align*}
H_{Kcq} = -\frac{K_{cq}}{2}a^{\dag 2} a^2 Z.
\end{align*}
For both these interactions we assume that they are only present during the $R(-Z\pi/2)$ gates (which is roughly 2/3 of the total time duration, assuming measument takes 1/3 of the time).

\subsubsection{Stochastic Errors}
The unitary operation generated by $H=\chi a^{\dagger} a Z+ H_{Kc}+H_{Kcq}$ is simulated using a Lindblad master equation for the qubit-cavity system for the duration of the qubit-cavity gates. In this Lindblad master equation we can also include stochastic sources such as photon loss from the cavity and amplitude damping for the qubit. We do not consider qubit dephasing. We will thus simulate the dynamics of a Lindblad equation of the form $\dot{\rho}=-i [H, \rho]+{\cal D}(\sqrt{\kappa}a)\rho+{\cal D}(\sqrt{\gamma}\sigma_-)  \rho$ with the compactly-defined superoperator ${\cal D}(X) \rho= X \rho X^{\dagger}-\frac{1}{2}(X^{\dagger} X\rho+\rho X^{\dagger} X)$. Due to limitations in the accuracy of the employed master equation solver (Python QuTip package), the study of these types of stochastic errors is limited to $M=8$ rounds (see \ref{sec:accuracy}).
Note that stochastic errors also play a role during qubit measurement.
% BMT Daniel i inserted Python, please check, it is good to make clear what you use, will the software be available on GitHub, if so add a link.
% DW QuTip is correct, I can publish most of the code (everything except the paralellization), will work on it when I'm done here

\subsubsection{Measurement Errors}
We will model two types of measurement errors: imperfect projection and readout errors. A {\em readout error} refers to the scenario in which a qubit is projected onto a state $\ket{x}$ and we learn $x$ with probability $1-p$ but $\bar{x}$ with probability $x$. Let $\Pi_x=\ketbra{x}{x}$, i.e. the projector onto qubit in state $\ket{x}$, $x\in \{0,1\}$. Our lack of information about the qubit outcome can be modeled as the following map applied to the qubit-cavity density matrix $\rho$ per round
\begin{align*}
&\rho \to (1-p)\Pi_x \rho \Pi_x + p\Pi_{\bar{x}} \rho \Pi_{\bar{x}}.
\end{align*}
{\em Imperfect projection} of the qubit refers to the scenario where a measurement of the ancilla qubit leads to us learning result $x$ while the cavity-qubit system undergoes the map 
\begin{align*}
&\rho \to A_x \rho A_x^{\dagger}, && A_x = \sqrt{1-p}\Pi_x + \sqrt{p}\Pi_{\bar{x}}.
\end{align*}
These models are not identical in that there are coherent cross error terms of the form $\Pi_x \rho \Pi_{\bar{x}}$ in the latter model while these are absent in the first. Note that both these errors not only affect the current round, but also change the qubit input state of the next measurement if the ancilla is reset by a $X$ flip depending on the measurement outcome.

% requires two periods of controlled-rotations by the interaction $\chi Z a^{\dagger} a$, each taking time $t=\frac{\pi}{2\chi}$, (interspersed by fast cavity displacement and a single qubit $X$) followed by single qubit-rotation and qubit measurement. 

%\subsection{Reset of the Ancilla Qubit}
%There are several possibilities to reset the ancilla qubit. One is to cool it to its ground state. This method improves the behavior under measurement errors, but is expected to be relatively slow \cite{}. The other options are to reset the ancilla with a Pauli $X$ gate conditioned on the measurement outcome (used in this work) or to track the frame and compensate by changing the feedback of the next round. The latter method is equivalent to the former in the case of noiseless Pauli gates, but should be preferred in an experimental setting.

\subsection{Simulation Results}

\begin{figure}[htb]
\includegraphics[width=\hsize]{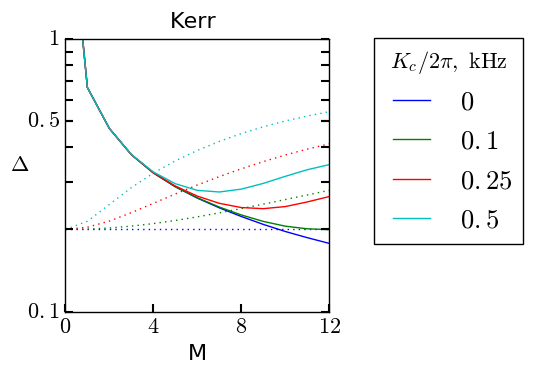}
\includegraphics[width=\hsize]{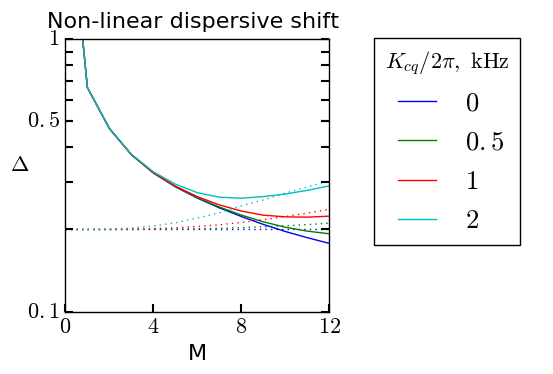}
\caption{Effective squeezing after M rounds, starting with a squeezed state, $\Delta=0.2$. Solid Lines: $\Delta_p$. Dashed lines: $\Delta_q$. Top: $H = \chi a^{\dagger}a Z+ H_{Kc}$, Bottom: $H =\chi a^{\dagger} a Z + H_{Kcq}$.}
\label{fig:K_single}
\end{figure}

\begin{figure}[htb]
\includegraphics[width=\hsize]{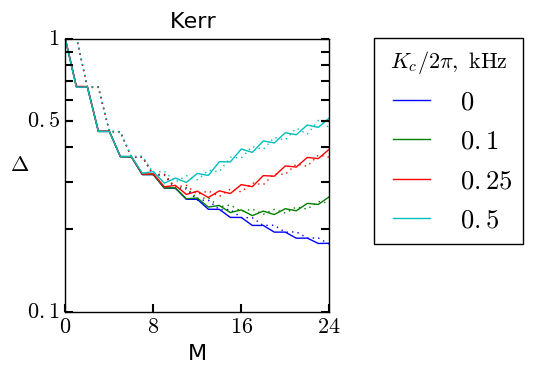}
\caption{Effective squeezing after M rounds of interleaved phase estimation with Kerr interaction, starting with the vacuum state. Solid Lines: $\Delta_p$. Dashed lines: $S_q$. Odd round numbers represent a measurement of $S_p$, even rounds a measurement of $S_q$.}
\label{fig:K_iter}
\end{figure}

\begin{figure}[htb]
\includegraphics[width=\hsize]{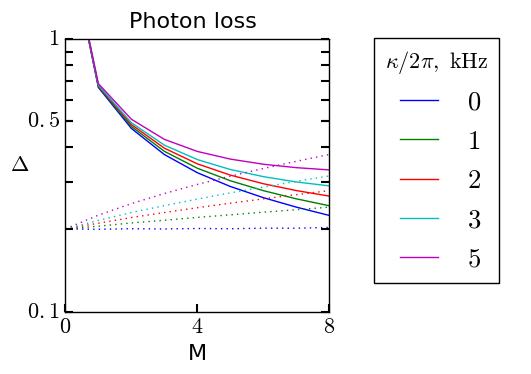}
\includegraphics[width=\hsize]{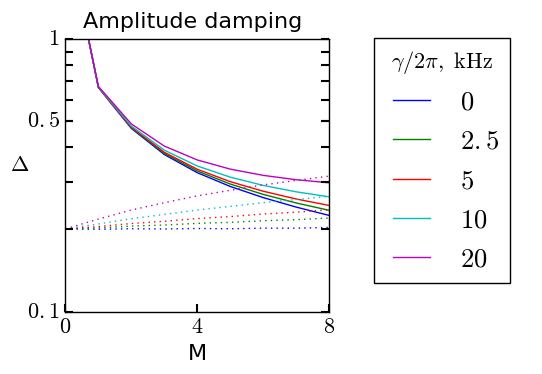}
\caption{Effective squeezing after M rounds, starting with a squeezed state, $\Delta=0.2$. Solid Lines: $\Delta_p$. Dashed lines : $\Delta_q$. Top: Photon loss, Bottom: Amplitude Damping.}
\label{fig:photon_loss}
\end{figure}

\subsubsection{Finite Hilbert Space}
\label{sec:accuracy}
The simulations are performed in the photon number basis, using a Hilbert space of finite size $N$. All operators acting on the system are obtained from truncated creation and annihilation operators. Thus, the simulation is only accurate if the support of a state outside of the finite Hilbert space is small. 
% BMT Daniel is the simulation time? or simulation space, why not O(N^3), how precise is this statement, if not precise please just say `A good comprise between comp. costs is the choice N=400 (ie. without making O(N) or O(N^2) statements
%DW Scaling: Memory N^2, Time: ODE (Dominant) at least N^2(stochastic, N^2 coupled ODEs) or N (else, equations are diagonal),
%Gates N^3 or N^2 depending on errors
%It is also possible to use e^(iHt) instead of the solver in the non-stochastic case, but the gains will not be big. 
%DW should be better now
As the simulation of non-stochastic errors is fairly efficient, we can choose $N=400$ for $M=12$ rounds. In case of stochastic errors acting over a finite time (e.g. photon loss), the accuracy of the simulation is also limited by the employed differential equation solver. As the precision is mostly independent of the size of the Hilbert space, the cumulative impact grows with its size. A suitable compromise between errors from the cutoff and errors from the ODE Solver is a Hilbert space with $N=100$ photons, allowing for $M=8$ rounds. While the computational cost for stochastic errors scales much less favorable than in the other cases, this would only be an issue for $N>100$.\\
% BMT clarify last sentence, it is not clear

\begin{figure}[htb]
\includegraphics[width=\hsize]{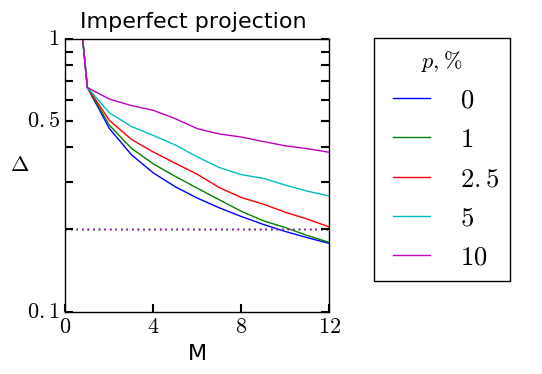}
\includegraphics[width=\hsize]{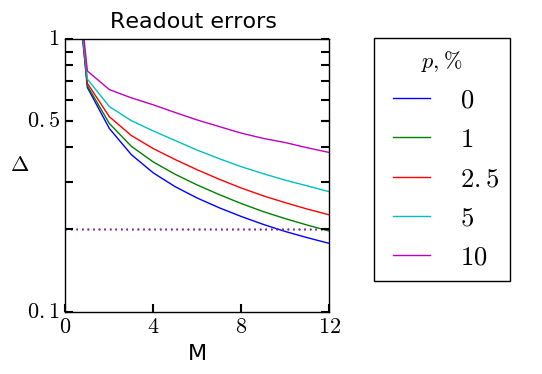}
\caption{Effective squeezing after M rounds, starting with a squeezed state, $\Delta=0.2$. Solid Lines: $\Delta_p$. Dashed lines: $\Delta_q$. Top: Imperfect projection, Bottom: Readout Errors.}
\label{fig:Measurement_errors}
\end{figure}

When simulating phase estimation for $S_p$ for $M$ rounds, all possible $2^M$ measurement results are simulated, each giving rise to a state $\rho({\bf x}_M)$. We then calculate the average squeezing parameter $\langle \Delta_p\rangle= \sum_M {\mathbb P}({\bf x}_M) \Delta_p(\rho({\bf x}_M))$, this is labeled as $\Delta$ on the vertical axis in Figs.~\ref{fig:K_single}-\ref{fig:Measurement_errors}. Note that this average does not depend on the value of $\tilde{\theta}_M=\arg {\rm Tr} S_p \rho({\bf x}_M)$.

% DW fixed $\Delta_p$ in squeezed vacuum, added value for $\Delta$
We start each simulation with a squeezed vacuum state in $q$ with $\Delta=0.2$, hence $\Delta_q=\Delta$ and $\Delta_p=1/\Delta$. By performing the $M$-round phase estimation circuits for $S_p$, $\Delta_p$ is gradually shrinking, but $\Delta_q$ will be gradually increasing due to photon loss and nonlinearities if no phase estimation measurements for $S_q$ are performed. This decrease in $S_p$ and increase in $\Delta_q$ is visible in all the data plots, Fig.~\ref{fig:K_single}-\ref{fig:photon_loss}. One can thus roughly take the increase in $\Delta_q$ as a measure of how the grid state deteriorates passively under photon loss or nonlinearities in time. 

%DW changed numbers to better fit figures
What is noticeable is that in the presence of nonlinearities, say the Kerr nonlinearity, $\Delta_p$ starts {\em increasing} after a certain number of rounds, implying that applying more rounds of phase estimation in fact decreases the quality of the state. Already a Kerr interaction of the order $K_c/2\pi = 500\mathrm{Hz}$ is sufficient to limit the procedure to $4$ rounds. 
For the non-linear anharmonicity, the situation is slightly better, i.e. $K_{cq}/2\pi = 2\mathrm{kHz}$ can be tolerated.
As the typical rates for both effects in an experimental setting are $\mathcal{O}(1)-\mathcal{O}(10 )\mathrm{kHz}$, treating them as systematic errors and correcting for them \cite{ofek+:QEC} is very relevant.

This effect is clearly not present for photon loss and amplitude damping of the qubit, Fig.~\ref{fig:photon_loss}, where the quality of the state gets better with the number of rounds. The highest amplitude damping rate of $20$kHz corresponds to a reasonable qubit decay time of $50\mu$sec. The effect of amplitude damping is small since qubit and cavity are only coupled for a short amount of time per round, $O(200)$nanosec. However note that when the qubit does jump from $\ket{1}$ to $\ket{0}$, the state will suffer a large stochastic displacement.

The robustness against photon loss, amplitude damping and measurement errors also suggests that a wider range of experimental settings could be explored for this type of encoding. For example, photon loss rates of the order $5\mathrm{kHz}$ are achievable in 2D microwave cavities \cite{BDD15}.

Since $S_p$ and $S_q$ commute, one can alternate or interleave the single round circuits in Fig.~\ref{fig:SPE} for $S_p$ and $S_q$, so that both $\Delta_p$ and $\Delta_q$ decrease or remain low. Since we actually use the circuits in Fig.~\ref{fig:PE_GKP} which contains a qubit-independent pre-displacement, e.g. $S_p^{-1/2}$, which does not commute with $S_q$, one needs to correct for these additional displacements when estimating the phase $\theta$.  Since the number of possible results is squared for alternating measurements, it is no longer possible to simulate all possible outcomes. Instead, the measurement process of the ancilla qubit is simulated, i.e. we take a total of $2000$ samples from the distribution of outcomes ${\mathbb P}({\bf x}_M)$. Afterwards, the results are weighed and averaged as before.

% BMT Daniel can you shorten and clarify the last 2 sentences here it is written in such a sketchy way
%DW done

In Fig.~\ref{fig:K_iter} we show how interleaving the measurement of $S_p$ and $S_q$, starting from the vacuum state leads to a grid state. Without the Kerr effect the quality of the state improves with the number of measurement rounds. Including the Kerr effect gives rise to an optimal number of rounds.
%Perhaps surprisingly, the qualitative behavior of the interleaved procedure is very similar to a measurement of $S_p$ only, although the time-scale is doubled in that case.
 Essentially, we expect that for a larger number of rounds (which means effectively a larger number of photons in the state) the Kerr effect introduces larger errors (see also the discussion in \cite{TW:GKP}). It is not visible from this numerical data whether or when the same saturation occurs in the presence of photon loss.

% performing the phase estimation measurement first for one of them and then the other. If the individual rounds of the measurements commute, it is also possible to interleave the measurements. An example of this type is discussed in sec \ref{sec:alternating}.
%If the rounds of the phase estimation procedures of two different operators commute, it is possible to interleave them. In the case of commuting operators, e.g. the stabilizers $S_p$ and $S_q$, this can be achieved by adding a displacement $D((-1)^m \sqrt{\pi/2})$ (for $S_q$: $D((-1)^m i\sqrt{\pi/2})$) after  the $m$-th round of $S_p$ ($S_q$). This is equivalent to replacing the gate $D(-\alpha/2)$ in fig. \ref{fig:PE_GKP} with the identity for $m$ even and $D(-\alpha)$ for $m$ odd.

The preparation protocol by multiple consecutive rounds of phase estimation is inherently robust against measurement errors. This is because all measurements contain some information of the whole eigenvalue distribution of the target state. Thus, if a single measurement result is flipped, it will be overridden after some number of additional measurements. We in fact expect that this form of phase estimation is much more robust to read-out noise than textbook phase estimation and could thus more generally be a preferred phase estimation protocol on partially-coherent qubits.

The preparation protocol is robust against photon loss, amplitude damping and readout errors, even large rates do not prohibit the generation of grid states, while additional measurements always improve the state in the cavity (for at least $M=12 (8)$ rounds). Some robustness against photon loss can be understood by expanding low-strength photon loss in terms of small displacements (see \cite{TW:GKP}).

%BMT Daniel what do you mean here??, time-scale is doubled, horizontal axes are the same in plots
%DW fixed in figure

%It remains to numerically simulate the performance of such circuit-QED protocol including decoherence and noise to estimate the practical sensitivity of the sensor for weak voltage signals. One expects that the sensitivity will depend on the strength of the electric field, the capacitive coupling between cavity and field, as well as the effective squeezing parameter $\Delta$ of the prepared sensor state.

%The protocols in \cite{TW:GKP} are not directly fault-tolerant with respect to qubit errors, but ideas can be formulated to make them more robust. 

 \section{Conclusion}

%The principal idea in this paper could be applied more widely, for example 
%That is, take two commuting Hermitian observable $A$ and $B$ on a $d$-dimensional space where the commutator is proportional to $iI$ such that $S_A=\exp(i \sqrt{2\pi}A)$ and $S_B=\exp(i \sqrt{2\pi}B)$ commute. Deviations from the state for which $S_A$ and $S_B$ have $+1$ eigenvalues can be measured by phase estimation for the operators $S_A$ and $S_B$. An example would be particle number $A=\hat{n}$ and properly-defined phase operator $B=\psi$.

%The idea applies quite generally. For example, we could choose a sensing state which is a simultaneous eigenstate of any even photon number and phase being $\phi=0 \mod \pi$ by taking $S_n=\exp(i \pi \hat{n})$ and $S_{\phi}= \exp( i 2\hat{\phi})$ with $[\hat{n},\hat{\phi}]=i$.
%Such state should allow one to determine XX $\Delta n \Delta \phi \geq 1/2$. Optical field phase.

We have presented the idea of a grid state sensor and shown how well this state could perform at a theoretical level as a displacement sensor which senses both real parameters of the displacement. We have presented some numerical data which show that one can prepare a grid state even using phase estimation with errors, confirming that nonlinearities can badly affect the quality of the state. 

If it is possible to prepare a grid state with some $\Delta_p$ and $\Delta_q$ and apply small displacements (adding at most $\pi/2$ photons), one could compare the performance of such grid state to a squeezed or coherent state in a simple metrology by repetition experiment. Such an experiment may be easier to implement than multi-round phase estimation.  In each round of the experiment one prepares the grid state with a certain $\Delta_p,\Delta_q$ (and estimates for the eigenvalues of $S_p$ and $S_q$). Then one lets an unknown small displacement act and then one runs a single Ramsey phase estimation circuit Fig.~\ref{fig:SPE} (taking $l=1$) for both $S_p$ {\em and} $S_q$.  By repeating the experiment (and assuming identical set-up) and plotting the oscillating probability for qubit outcome 0, one infers the shifted eigenvalues for $S_p$ and $S_q$ due to the displacement, hence learning both parameters of the displacement with accuracy limited by $\Delta_p$ and $\Delta_q$ (and the number of rounds of repetition). It is then also possible to use different feedback phases in each experiment to gain information more efficiently as in the phase estimation in Section \ref{sec:choice}.

Running the same experiment with a squeezed state input (with the same $\Delta$ as the grid state) would let one figure out one of the displacement parameters with similar accuracy (picked up by, say, $S_p$), but the Ramsey signal would be much more noisy for the $S_q$ measurement. For a coherent state, say the vacuum state, both Ramsey signals would have unsqueezed Gaussian noise due to ${\rm Var}(q)={\rm Var}(p)=1/2$.

BMT would like to thank Seth Lloyd, Olivier Pfister and David DiVincenzo for interesting discussions. This work is supported by the European Research Council (EQEC, ERC Consolidator Grant No: 682726).

%\bibliography{cat_refs}
\input{displace_sensor16.bbl}
\appendix 

\section{Two-Parameter Quantum Cram\'er-Rao bound}
\label{sec:2para}

We provide a self-contained derivation of the two-parameter quantum Cram\'er-Rao bound which, in slightly different form, was originally derived in \cite{helstrom:quantum_CR}.  We then consider this bound for the problem of displacement sensing using a pure state. We start by deriving a weaker bound 
in Eq.~(\ref{eq:var}) for which we discuss the reasons why it will not be tight. Then we derive the matrix inequality quantum Cram\'er-Rao bound which is stated and used in the main text.

%The optimal quantum measurement achieving the quantum Cramer-Rao bound will depend on the quantum state, i.e. it is determined by the derivatives of the parametrized state with respect to the parameters. This measurement thus aims to capture in what direction the state is moving. This is different from a model in which we assume that the state has been displaced by a small finite amount and we seek to determine by how much exactly. 

Let $\theta_1$ and $\theta_2$ be two (real) parameters and let ${\mathbb P}(x|\theta_1, \theta_2)$ be the probability  distribution over the data $x$
given the two parameters. Given a quantum state wich depends on the parameters $\rho_{\theta_1,\theta_2}$, ${\mathbb P}(x|\theta_1, \theta_2)$ may be obtained through a quantum measurement with outcomes $x$, i.e. ${\mathbb P}(x|\theta_1, \theta_2)={\rm Tr}_S E_x \rho_{\theta_1,\theta_2}$, $\sum_x E_x=I$ (and ${\rm Tr}_S$ is just the trace over the system Hilbert space). 
We will be interested in bounding the variance of any locally unbiased estimator $\tilde{\theta}_i(x)$ for $\theta_i$ around the point $(\theta_1,\theta_2)$. For such a locally unbiased estimator (hence MSD equals Variance) at $(\theta_1,\theta_2)$ it holds that 
\begin{equation}
\forall, i,j,\;\frac{\partial}{\partial \theta_i}\left(\sum_x {\mathbb P}(x|\theta_1, \theta_2) (\tilde{\theta}_j(x)-\theta_j)\right)=0, \nonumber
\end{equation}
or
\begin{equation}
\sum_x  \frac{\partial {\mathbb P}(x|\theta_1, \theta_2)}{\partial \theta_i}(\tilde{\theta_j}(x)-\theta_j)=\delta_{ij}. 
\nonumber
\end{equation}
One can rewrite the last equation as 
\begin{equation}
\sum_x {\rm Re}({\rm Tr}_S( \rho_{\theta_1,\theta_2} E_x L_{\theta_i}))\left(\tilde{\theta_j}(x)-\theta_j\right)=\delta_{ij}, \label{eq:unb}
\end{equation}
where the Hermitian symmetric logarithmic derivative operator $L_{\theta_i}$ is defined by
\begin{equation}
\frac{\partial \rho_{\theta_1,\theta_2}}{\partial \theta_i}=\frac{1}{2}\left( L_{\theta_i} \rho_{\theta_1,\theta_2}+\rho_{\theta_1,\theta_2} L_{\theta_i}\right). \nonumber  
\end{equation}
We formally define
\begin{eqnarray}
P=\sum_{x,i} E_x^{1/2} L_{\theta_i} \rho_{\theta_1,\theta_2}^{1/2} \otimes \ketbra{x}{x}_A \otimes \ketbra{i}{i}_B, \nonumber \\
Q=\sum_{x,i}  E_x^{1/2}\rho_{\theta_1,\theta_2}^{1/2}   \left(\tilde{\theta_i}(x)-\theta_i\right) \otimes \ketbra{x}{x}_A \otimes \ketbra{i}{i}_B, \nonumber 
\end{eqnarray}
where the new Hilbert spaces $A$ and $B$ are defined for notational convenience. We can use the Cauchy-Schwarz inequality and Eq.~(\ref{eq:unb}) so that $4=({\rm Re}({\rm Tr}_{SAB}\, Q^{\dagger} P))^2\underset{(I)}{\leq} |{\rm Tr}_{SAB}\,(Q^{\dagger} P)|^2 \underset{(II)}{\leq} {\rm Tr}_{SAB}\, (Q^{\dagger} Q) {\rm Tr}_{SAB}\,(P^{\dagger} P)$ so that ${\rm Tr}_{SAB}\, (Q^{\dagger} Q) =$
\begin{equation}
{\rm Var}(\tilde{\theta}_1)+{\rm Var}(\tilde{\theta}_2) \geq \frac{4}{{\rm Tr}(F)},
\label{eq:var}
\end{equation}
where the quantum Fisher information matrix operator of a state $\rho_{\theta_1,\theta_2}$ is $F_{ij}=\frac{1}{2} {\rm Tr}( \rho_{\theta_1,\theta_2} (L_{\theta_i} L_{\theta_j}+L_{\theta_j} L_{\theta_i}))$. Here ${\rm Tr}(F)=\sum_i F_{ii}$.  

One can examine whether one can achieve this bound in the single or multi-parameter case; this depends on whether inequalities $(I)$ and $(II)$ are tight. 
However, as is well known, a measurement which achieves this bound typically depends on the parameter to be estimated and is thus not of immediate interest (see \cite{BCM:uncertainty} for optimal parameter-value independent measurement in the single-parameter case with pure states). 
Here we briefly discuss why it is hard to achieve the bound for the two-parameter displacement estimation problem on pure states.

The Schwarz inequality $(II)$ is tight when we have 
\begin{equation}
\exists \mu, \forall x,i\;\colon E_x^{1/2} \rho_{\theta_1,\theta_2}^{1/2} (\tilde{\theta}_i(x)-\theta_i)=\mu E_x^{1/2} L_{\theta_i} \rho_{\theta_1,\theta_2}^{1/2}.
\label{eq:sat}
\end{equation}
We restrict ourselves to pure states $\rho_{\theta_1,\theta_2}=\ketbra{\psi}{\psi}$ where $\psi=\psi(\theta_1,\theta_2)$. For a pure state it can be shown that $L_{\theta_i}=2i[K_{\theta_i},\ketbra{\psi}{\psi}]$ with hermitian $K_{\theta_i}$. Let us first re-examine tightness in the single parameter case, as it was first discussed in \cite{BC:stat_dist}. One can choose a measurement $E_x$ which is a projective measurement in the eigenbasis of $L_{\theta}$. This choice then determines the choice for $\tilde{\theta}(x)$ as a function of $x$ and the eigenvalues of $L_{\theta}$. As an example let $2 i K_{\theta} \ket{\psi}=\ket{d\psi_K}$ and assume that we work at a point $\psi(\theta)$ where $\bra{d\psi_K} \psi\rangle \propto \bra{\psi} K \ket{\psi}=0$. We have 
\begin{equation}
L_{\theta}=dX_K (\ketbra{+_K}{+_K}-\ketbra{-_K}{-_K}) \nonumber,
\end{equation}
where $dX_K=\sqrt{\bra{d\psi_K}d\psi_K\rangle}$ and $\ket{\pm_K}=\frac{1}{\sqrt{2}}(\ket{\psi}\pm \ket{\psi_K})$ with $\ket{\psi_K}=\ket{d\psi_K}/\sqrt{\bra{d\psi_K}d\psi_K\rangle}$. The measurement $\{E_x\}$ in the eigenbasis of $L_K$ is a measurement in the orthogonal basis $\ket{\pm_K}$ and states spanning the space orthogonal to $\ket{\pm_K}$ (the latter have zero probability given that the infinitesimal deviation of the state is only in the direction of $\ket{\psi_K}$). Upon outcomes $x=\pm$, Eq.~(\ref{eq:sat}) constrains how to choose $\tilde{\theta}(\pm)$, i.e.
\begin{equation}
\exists \mu, \tilde{\theta}(\pm)-\theta=\pm \mu dX_K. \nonumber
\end{equation}
Note that this last equality does not fix the unbiased estimator due to the freedom of $\mu$. With such a choice of measurement, the first inequality, $(I)$, is also satisfied since all quantities involved are real.

In the two-parameter case the operators $L_{\theta_1}$ and $L_{\theta_2}$ may not be commuting implying that a basis in which both are diagonalized does not exist.  Let us in particular consider the displacement case for pure states to understand what goes wrong.  In this case we have $K_{\theta_1}=-\hat{p}$ and $K_{\theta_2}=\hat{q}$. For simplicity, we consider displacement deviations of a pure state $\ket{\psi}$ with the property that $\bra{\psi} \hat{p}\ket{\psi}=\bra{\psi} \hat{q}\ket{\psi}=0$ and furthermore $\bra{\psi} \hat{p} \hat{q} \ket{\psi}=-i/2$ (as $\bra{\psi} \hat{p}\hat{q} \ket{\psi}=-\bra{\psi} \hat{q}\hat{p}\ket{\psi}$). These properties hold for the photon number state, grid state and quantum compass state in this paper (when these states are symmetrically centered around the vacuum state in phase phase). In general the expectation $\bra{\psi} \hat{p} \hat{q} \ket{\psi}$ equals $-i/2$ plus the expectation value of a Hermitian operator and can thus never be real!
 
Let $\ket{\psi_p}=\ket{d\psi_p}/\sqrt{\bra{d\psi_p}d\psi_p\rangle}$ and similarly $\ket{\psi_q}$. For the states considered we have $dX=dX_p=\sqrt{\bra{d\psi_p}d\psi_p\rangle}=\sqrt{\bra{d\psi_q}d\psi_q\rangle}=\sqrt{4 \overline{n}+2}$, showing that a small displacement leads to a change in the state which grows with photon number. This in turn implies that $\bra{\psi_p} \psi_q\rangle=\frac{-i}{2\overline{n}+1}$.
{\em If} $\ket{\psi_p}$ and $\ket{\psi_q}$ were orthogonal it can in fact be shown that one can construct a rank-1 four-outcome POVM measurement which saturates $I$ and $II$. However, for the displacement problem $\psi_p$ and $\psi_q$ are non-orthogonal, although their innerproduct vanishes as $1/\overline{n}$. Implicitly, we see that non-commutativity of $p$ and $q$ implies that $\bra{\psi} \hat{p} \hat{q} \ket{\psi}$ has a nonvanishing imaginary part which provides an obstruction for satisfying $I$ and $II$. 

%This suggests that one may be able to construct a POVM measurement and estimators such that ${\rm Var}(\tilde{\theta}_1)+{\rm Var}(\tilde{\theta}_2) \geq \frac{c}{\sqrt{n}}$. 

\subsubsection{Matrix Inequality}
We derive the more stringent matrix version of the two-parameter quantum Cram\'er-Rao bound $\Sigma \geq F^{-1}$. The covariance matrix $\Sigma$ of the unbiased estimators $\tilde{\theta}_i(x)$ is defined as
\begin{equation}
\Sigma_{ij}=\sum_x {\mathbb P}(x|\theta_1,\theta_2) (\tilde{\theta}_i(x)-\theta_i)(\tilde{\theta}_j(x)-\theta_j), \nonumber
\end{equation}
so that ${\rm Var}(\tilde{\theta}_1)+{\rm Var}(\tilde{\theta}_2)={\rm Tr}(\Sigma)$. Define $L(a)=\sum_i a_i L_{\theta_i}$ and $T(b)=\sum_i b_i  (\tilde{\theta_i}(x)-\theta_i)$ and operators 
\begin{eqnarray}
P_a=\sum_x E_x^{1/2} L(a) \rho_{\theta_1,\theta_2}^{1/2} \otimes \ketbra{x}{x}_A, \nonumber \\
Q_b=\sum_x  E_x^{1/2}\rho_{\theta_1,\theta_2}^{1/2} T(b)  \otimes \ketbra{x}{x}_A. \nonumber 
\end{eqnarray}
Then $\bra{b} a\rangle={\rm Re}\left({\rm Tr}_{SA} (Q_b^{\dagger} P_a)\right)$ where we have included the $\sum_x$ as a trace over the ancillary space $A$ for convenience. Now one uses the Cauchy-Schwarz inequality $|{\rm Tr}\,(Q_b^{\dagger} P_a)|^2 \leq {\rm Tr}\, (Q_b^{\dagger} Q_b) {\rm Tr}\,(P_a^{\dagger} P_a)$.

We can write 
\begin{equation}
{\rm Tr}\, (Q_b Q^{\dagger}_b) =\bra{b} \Sigma \ket{b}, \; {\rm Tr}\,(P_a^{\dagger} P_a)=\bra{a} F \ket{a}. \nonumber
\end{equation}
We can thus derive the inequality $|\bra{a} b\rangle|^2 \leq \bra{b} \Sigma \ket{b} \bra{a} F \ket{a}$ for all vectors $a,b$, or
\begin{equation}
\forall b,\;\bra{b} \Sigma \ket{b} \geq \max_{a} \frac{|\bra{a} F^{1/2} F^{-1/2} |b\rangle|^2}{\bra{a} F \ket{a}}= \bra{b} F^{-1}\ket{b}, \nonumber
\end{equation}
by taking $\ket{a}= F^{-1} \ket{b}$ and normalized vector $\ket{b}$, thus leading to $\Sigma \geq F^{-1}$. From $\Sigma \geq F^{-1}$ it follows that ${\rm Var}(\tilde{\theta}_1)+{\rm Var}(\tilde{\theta}_2) \geq {\rm Tr}(F^{-1})$. Note that 
${\rm Tr}(F^{-1})=\frac{{\rm Tr} F}{\lambda_0\lambda_1} \geq \frac{4}{{\rm Tr}F}$ where $\lambda_0,\lambda_1$ are the eigenvalues of $F$, hence Eq.~(\ref{eq:var}) is weaker than the inequality derived in this section.

%The matrix inequality can clearly not be achieved in general

\section{Analysis of Textbook Phase Estimation on Grid State} 
\label{sec:phase}

%DW QUestion: This figure has a large overlap with figs 2/4. Remove this one and refer to them instead?
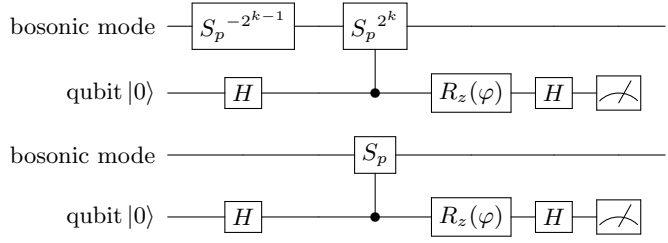
\begin{figure*}[h!tb]
\centering
\parbox{150pt}{
\Qcircuit @C=1em @R=1em {
\lstick{\mbox{bosonic mode}} & \gate{{S_p}^{-2^{k-1}}} & \qw & \gate{{S_p}^{2^k}} & \qw & \qw & \qw & \qw \\
\lstick{\mbox{qubit} \ket{0}} & \gate{H} & \qw & \ctrl{-1} &  \gate{R_z(\varphi)} & \gate{H} & \meter \\
\lstick{\mbox{bosonic mode}} & \qw & \qw & \gate{{S_p}} & \qw & \qw & \qw & \qw \\
\lstick{\mbox{qubit} \ket{0}} & \gate{H} & \qw & \ctrl{-1} &  \gate{R_z(\varphi)} & \gate{H} & \meter
}}
\caption{Circuits used in the sequential implemention of textbook phase estimation for the unitary operator $S_p$. This sequential realization of phase estimation is identical to normal phase estimation as it merely uses a semi-classical realization of the quantum Fourier transform so that a single qubit is used in each round (see for example \cite{ME:hidden}). Assume that phase estimation uses $M$ ancillas prepared in $\ket{00 \ldots 0}$. The top circuit in this Figure is repeatedly executed for $k=M-1, \ldots, 1$ starting at $k=M-1$, each circuit using one qubit. For $k=0$ we use the circuit at the bottom, i.e. we do not 
apply the unconditional displacement $S_p^{-1/2}$: this has no effect on the phase estimation protocol, but ensures that the measurement for $S_p$ and $S_q$ fully commute. The only reason to have the unconditional displacement in the top circuit is that one minimizes the overall number of photons used in the protocol by doing so. The phase $\varphi$ in $R_z(\varphi)$ will depend on the outcomes of all previously measured qubits in accordance with the description of the semi-classical Fourier transform.}
\label{fig:SPE_part}
\end{figure*}

\begin{figure*}[htb]
\centering
\parbox{100pt}{
\Qcircuit @C=1em @R=1em {
&\lstick{\ket{\rm vac.}} & \gate{\rm S_p\; Meas \rightarrow x_p} & \gate{\rm S_q\;Meas\rightarrow x_q} & \gate{e^{i v\hat{q}}} & \gate{e^{-i u \hat{p}}} & \gate{\rm S_p\;Meas \rightarrow x'_p} & \gate{\rm S_q\; Meas \rightarrow x'_q}\qw & \qw\\
= \\
&\lstick{\ket{\rm vac.}} & \gate{\rm S_p\; Meas \rightarrow x_p} & \gate{e^{i v\hat{q}}}  & \gate{\rm S_p\;Meas \rightarrow x'_p} & \gate{\rm S_q\;Meas\rightarrow x_q}   & \gate{e^{-i u \hat{p}}} & \gate{\rm S_q\; Meas \rightarrow x'_q}\qw & \qw\\
}}
\caption{The unit $S_p$ Meas (resp. $S_q$ Meas) with outcomes $x_p$ or $x_p'$ (resp. $x_q$ and $x_q'$) represents the approximate measurement of the eigenvalue of $S_p$ (resp. $S_q$) by repeatedly using the circuits in Fig.~\ref{fig:SPE_part}. The circuit identity follows from the fact that the $S_p$ measurement and the $S_q$ measurement commute as (powers of) $S_p$ and $S_q$ commute.}
\label{fig:sequence}
\end{figure*}
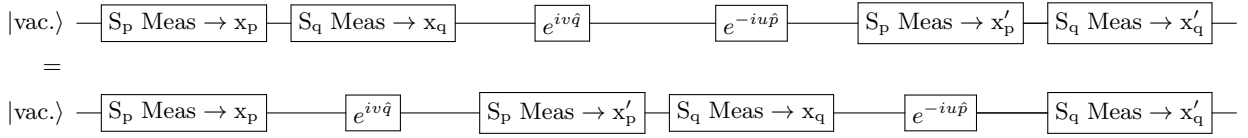

Let the preparation of the grid state starting from the vacuum state result in $M$-bit strings $x_p$ and $x_q$ for the phase estimation measurement of $S_p$ and $S_q$ respectively, leading to a pure state $\ket{\psi_{\rm grid}^{x_p,x_q}}$. Then an unknown displacement $D(\beta)=\exp(-iu \hat{p}+i v\hat{q})$ occurs where $u,v$ are assumed to be uniformly distributed in the interval $u,v \in  [-\sqrt{\pi/2},\sqrt{\pi/2})$. We then perform phase estimation measurements of $S_p$ and $S_q$ again using $M$ ancillas, giving output bitstrings $x'_p, x'_q$. The circuits used are the ones in Figs.~\ref{fig:SPE_part},\ref{fig:sequence}. The accessible information of this protocol is the mutual information between the variables $u, v$ and the measurement data $x_p, x_q, x'_q, x'_p$. Of course, the data $x_p,x_q$ do not depend on the displacement, but including this data and not fixing a particular input state allows us to include the possible variation in quality of the input state (in this analysis of phase estimation it turns out not to matter).

If the preparation of the state and the measurement are perfect quantum circuits, the process of gaining information about $u$ is identical and independent of the process of gaining information about $v$. One can commute through operators as in Fig.~\ref{fig:sequence} so that one first prepares an approximate eigenstate of $S_p$, the displacement $\exp(iv\hat{q})$ acts and one approximately measures the eigenvalue of $S_p$. After this step the same process with displacement $\exp(-i u\hat{p})$ happens for $S_q$ \footnote{If we pre-displace the bosonic mode by $S_p^{-1/2}$ in the $k=0$ circuit for $S_p$, see Fig.~\ref{fig:SPE_part} (and similarly for $S_q$), the $S_p$ and $S_q$ circuits would not commute, but the commutation would simply induce an additional rotation on a qubit since the gate $S_p^{-1/2}$ {\em does not commute} with the qubit-controlled $S_q$ gate and similarly for the circuit of phase estimation for $S_q$.}.  

Thus the accessible information is the sum of the mutual information between $u$ and $x_q,x'_q$ and the mutual information between $v$ and $x_p,x'_p$.
Since these two contributions are identical, we can take one process, say, for $S_p$ and displacement $\exp(i v\hat{q})$ and compute the mutual information for that one.
 We will drop the label $p$ from $x$ and $x'$ for notational convenience from now on. 
 
Let $\tilde{v}$ be the estimate for $v$, given the data $x,x'$ and let ${\mathbb P}(\tilde{v}|v)$ the probability for $\tilde{v}$ given $v$. 
The mutual information $I(v:\tilde{v})=H(\tilde{v})-H(\tilde{v}|v)$, with Shannon entropy $H(\tilde{v})$, can be written as
\begin{eqnarray}
I(v:\tilde{v}) & = & \int_S dv \sum_{\tilde{v}} {\mathbb P}(\tilde{v}|v) \log \frac{{\mathbb P}(\tilde{v}|v)}{{\mathbb P}(\tilde{v})}, \label{eq:mi}
\end{eqnarray}
where $\int_S dv=\frac{1}{\sqrt{2\pi}} \int_{-\sqrt{\pi/2}}^{\sqrt{\pi/2}} dv$. Note that the estimates $\tilde{v}$ will take a discrete set of values while $v$ ranges in principle continuously over the allowed interval. In numerical evaluations of this information we replace the continuous integration by a discrete Riemann sum so that both variables take on discrete values: one then lets the discretization of $v$ become sufficiently fine so that the information is independent of it.

% kasper, mee eens met bovenstaande tekst? misschien niet helemaal nodig dit detail te vermelden...

% For numerical evaluations the average $\int_S dv$ will always read 
%The entropy of the source, namely $v$, is given by $H(v)=-\int_{-\sqrt{\pi/2}}^{\sqrt{\pi/2}} dv \;{\rm Prob}(v) \log({\rm Prob}(v))=\frac{1}{2} \log (2\pi)$ for uniformly distributed $v$. ${\rm Prob}(v)=\frac{1}{\sqrt{2\pi}}$. Don't use this

% BMT commuting through S_p and S_q etc does it give a phase shift on the qubit so phase estimation protocol should be changed to reflect this, not an issue

We determine ${\mathbb P}(\tilde{v}|v)$ assuming that the phase estimation protocol is executed using textbook phase estimation \cite{book:nielsen&chuang, cleve+:revisited}, using circuits of the form depicted in Fig.~\ref{fig:SPE_part}. 
% In section \ref{sec:pe_heisenberg} we briefly discuss the merits of performing phase estimation in different ways.
 Let $N=2^M$. We can capture the phase estimation circuits for $S_p$ interspersed with the displacement $\exp(i v q)$ as a total measurement operator applied to the vacuum input state, i.e. $M_{x,x'}^v \ket{\rm vac}$.This measurement operator is obtained as $M_{x,x'}^v=\bra{x,x'} V \ket{0^{\otimes{2M}}}$, where $V$ is the total unitary in the circuits in Fig.~\ref{fig:SPE_part} and $\ket{0^{\otimes 2M}}$ is the initial state of the $2M$ qubit ancillas (both for $S_p$ and $S_q$). It can be shown (see similar statements in \cite{TW:GKP}) that $M_{x,x'}^v$ equals
\begin{eqnarray}
M_{x,x'}^v=\frac{1}{N^2} \sum_{t,t'=0}^{N-1}e^{-\frac{2\pi  i}{N} (x t+x' t')} S_p^{t'-\frac{N}{2}+1} e^{i v \hat{q}} S_p^{t-\frac{N}{2}+1}\, \label{eq:defM}
\end{eqnarray}
Using $x,x'$ one obtains $\tilde{v}$. Hence we calculate the probability ${\mathbb P}(\tilde{v}|v)=\sum_{(x,x') \rightarrow \tilde{v}}{\rm Tr}M_{x,x'}^{v \dagger} M_{x,x'}^v \ketbra{\rm vac}{\rm vac}$ where we sum over all pairs of $M$-bit strings $(x,x')$ which let us infer a particular value $\tilde{v}$. Allowing for any value of $x$ means that we do not preselect the state that we use, in fact any approximate eigenstate of $S_p$ will do fine. Using $I=\int dp \ketbra{p}{p}$ to replace $S_p^t$ by its eigenvalues we have 
\begin{eqnarray}
\lefteqn{{\mathbb P}(\tilde{v}|v)= \sum_{(x,x')\rightarrow \tilde{v}} \int dp |\bra{p} M_{x,x'}^v \ket{{\rm vac}}|^2} \nonumber \\
 & & \mbox{with }|\bra{p} M_{x,x'}^v \ket{{\rm vac}}|=|\bra{p-v} {\rm vac}\rangle \alpha_p(x') \alpha_{p-v}(x)| \nonumber
 \end{eqnarray}
and
\begin{equation}
\alpha_p(x) = \frac{1}{N} \sum_{t=0}^{N-1} e^{-\frac{2\pi  i}{N}(xt) +ip\sqrt{2\pi}t} \nonumber.
  \end{equation}
 We can use $\bra{p-v} {\rm vac}\rangle=\frac{1}{\pi^{1/4}}e^{-(p-v)^2/2}$ and perform the integral over $p$ (using $\frac{1}{\sqrt{\pi}}\int dp \,e^{-p^2} e^{-2 \pi i s p}= e^{-\pi^2 s^2}$), to get
\begin{eqnarray}
{\mathbb P}(\tilde{v}|v)= \sum_{(x,x')\rightarrow \tilde{v}}\frac{1}{N^4}\sum_{t_1=0,\ldots,t_4=0}^{N-1}e^{\frac{2\pi i}{N} (x (t_2-t_1)+x'(t_4-t_3))} \nonumber \\
e^{-i v\sqrt{2\pi} (t_4-t_3)} e^{-\frac{\pi}{2}(t_2-t_1+t_4-t_3)^2}.\nonumber
\end{eqnarray}
 How does $x,x'\rightarrow \tilde{v}$ take place: this is essentially described by the phase estimation algorithm \cite{book:nielsen&chuang, cleve+:revisited}. 
 In the usual analysis in phase estimation, the goal is to get a $\tilde{M}$-bit estimate $\tilde{v}$ for $v$ where $M=\tilde{M}+\lceil \log_2 (2+\frac{1}{2\epsilon})\rceil$: the $\epsilon$-dependent overhead boosts the probability of success. Here, we will not use any $\epsilon$-dependent overhead and use all the bits $x$ and the bits $x'$ to get a $M$-bit estimate $\frac{\tilde{v}}{\sqrt{2\pi}}$ for $\frac{v}{\sqrt{2\pi}}\in [-1/2,1/2)$. We can represent the eigenvalue as $\exp(2 \pi i\phi)$ with $0 \leq \phi < 1$ so that the phase $\phi$ can be expanded in binary as $0.\phi_1\ldots \phi_M=\sum_{j=1}^{M} 2^{-j} \phi_j $.  We set the bits of the initial phase $\phi^{\rm in}_i=x_i$ (or $\phi^{\rm in}=x/N$) and the bits of the output phase as $\phi^{\rm out}_i=x_i'$. Then we take $\frac{\tilde{v}}{\sqrt{2\pi}}$ such that $\phi^{\rm out}(x')=\phi^{\rm in}(x)+\frac{\tilde{v}}{\sqrt{2\pi}}$ where we use the periodicity of the phases to choose a $\frac{\tilde{v}}{\sqrt{2\pi}}\in [-1/2,1/2)$. This implies that $\sum_{(x,x')\rightarrow \tilde{v}}$ ranges over all $x=0, \ldots, N-1$, and $x'$ is then uniquely fixed by $x$ and $\tilde{v}$. In other words we can write
 \begin{eqnarray}
{\mathbb P}(\tilde{v}|v)= \sum_{x=0}^{N-1}\frac{1}{N^4}\sum_{t_1=0,\ldots,t_4=0}^{N-1}e^{\frac{2\pi i x}{N}(t_2-t_1+t_4-t_3))}\nonumber \\
e^{i(\tilde{v}-v)\sqrt{2\pi}(t_4-t_3)}e^{-\frac{\pi}{2}(t_2-t_1+t_4-t_3)^2}. \nonumber
\end{eqnarray}
 We can now perform the Fourier sum $\sum_{x=0}^{N-1} \exp(\frac{2 \pi i x s }{N})=N \delta_{s0}$ with integer $s$, leading to the much simplified expression
  \begin{equation}
{\mathbb P}(\tilde{v}|v)= \left|\frac{1}{N}\sum_{t=0}^{N-1} e^{i(\tilde{v}-v)\sqrt{2\pi}t} \right|^2= \left|\frac{1}{N}\sum_{t=0}^{N-1} e^{2 \pi i  \delta(\tilde{v}-v)t} \right|^2, \nonumber
\end{equation}
 with $\delta(\tilde{v}-v)=\frac{1}{\sqrt{2\pi}}(\tilde{v}-v)$. One can rewrite this expression using the geometric series so that we get
 \begin{equation}
{\mathbb P}(\tilde{v}|v)=\frac{1}{N^2} \frac{\sin^2( \pi N \delta(\tilde{v}-v))}{\sin^2(\pi \delta(\tilde{v}-v))}.
\label{eq:expressP}
\end{equation} 
 By performing the integral $\int_S dv \,e^{-i v \sqrt{2\pi}(t-t')}=\delta_{tt'}$, one obtains, as expected, 
  %BMT for a Gaussian distrubution over v, this may become a Gaussian distribiution over tilde{v}, just integrate some Gaussian and phase factors
\begin{equation}
{\mathbb P}(\tilde{v})=\int_S dv\, {\mathbb P}(\tilde{v}|v)=\frac{1}{N}.\nonumber
\end{equation}
 We can thus  express the mutual information in Eq.~(\ref{eq:mi}) as
\begin{equation}
I(v:\tilde{v})=M +  \frac{1}{\sqrt{2\pi}} \int_{-\sqrt{\pi/2}}^{\sqrt{\pi/2}} dv \sum_{\tilde{v}} {\mathbb P}(\tilde{v}|v) \log {\mathbb P}(\tilde{v}|v), \nonumber
\end{equation}
where the last part will substract some of the information gained. One estimates $\frac{\tilde{v}}{\sqrt{2\pi}}=-\frac{1}{2}+\frac{y}{N}$ with $N$-bit integer $y \in \{0,1\}^N$, hence the sum over $\tilde{v}$ can be replaced by a sum over bit-strings $y$. It is most useful to numerically evaluate the mutual information $I_{\rm acc}(M)=I(v:\tilde{v})+I(u:\tilde{u})$ as a function of $2M$, see Fig.~\ref{fig:entropy_bound}.
 
% BMT on unbiased estimator and operator to be measured.

\subsubsection{Mean Squared Deviation and Variance of Estimates}

  The full measurement to estimate $v$ is described by the POVM elements $E(\tilde{v}|v)=\sum_{(x_p,x_p') \rightarrow \tilde{v}} P_{x_p,x'_p}^v$ with $P_{x_p,x'_p}^v= M_{x_p,x_p'}^{v \dagger} M_{x_p,x_p'}^{v} \otimes \ketbra{x_p,x'_p}{x_p,x'_p}$ so that $\sum_{\tilde{v}} E(\tilde{v}|v)=\sum_{x_p,x'_p} P_{x_p,x'_p}^v=I$, using Eq.~(\ref{eq:defM}). We have ${\rm Tr} (E(\tilde{v} |v)\ketbra{\rm vac}{\rm vac})=
\mathbb{P}(\tilde{v}|v)$. The estimator thus has expectation $\sum_{\tilde{v}} \tilde{v} \;\mathbb{P}(\tilde{v} |v)$.  
The mean squared deviation (MSD) which captures the real performance of the protocol is given by 
\begin{equation}
{\rm MSD}(\tilde{v})=\sum_{\tilde{v}}\mathbb{P}(\tilde{v} |v) (\tilde{v}-v)^2,
\end{equation}
and similarly ${\rm MSD}(\tilde{u})$. One has ${\rm MSD}(\tilde{v})={\rm Var}(\tilde{v})+b^2(\tilde{v})$ with bias $b(\tilde{v})=\sum_{\tilde{v}} \mathbb{P}(\tilde{v}|v) \tilde{v}-v$ (using the definition of the variance ${\rm Var}(\tilde{v})=\sum_{\tilde{v}}\mathbb{P}(\tilde{v} |v) (\tilde{v}-\sum_{\tilde{v}} \tilde{v} \;\mathbb{P}(\tilde{v} |v) )^2$).
Only in the interval when $\frac{v}{\sqrt{2\pi}} \in (-\frac{1}{2},\frac{1}{2}]$, we expect the mean squared deviation to be small as $\tilde{v}$ only takes values in this interval as well. In addition, our estimate $\tilde{v}$ will be only close to $v \mod \sqrt{\pi}$. i.e. we cannot distinguish $v=-\sqrt{\pi/2}$ from $v=\sqrt{\pi/2}$. There will be a non-neglible probability to pick, say, a $\tilde{v}/\sqrt{2\pi}$ close to $-1/2$ when $v$ is close to $1/2$ due to the periodicity of $\mathbb{P}(\tilde{v}|v)$. Hence one can only hope to get a good upper bound on the mean squared deviation of $\tilde{v}$ in a restricted interval $v\in I$ with $I=[-\sqrt{\frac{\pi}{2}}+2\alpha \sqrt{\frac{\pi}{2}},\sqrt{\frac{\pi}{2}}-2\alpha \sqrt{\frac{\pi}{2}}]$ with a constant $0<\alpha< \frac{1}{2}$. The mean squared deviation for different values of $v$ depends on $N$, i.e. we estimate $\frac{\tilde{v}}{\sqrt{2\pi}}=-\frac{1}{2}+\frac{y}{N}$ with $N$-bit integer $y \in \{0,1\}^N$.  The worst-case MSD is clearly obtained when $v$ falls right in the middle between two values for $\tilde{v}$, that is, $\frac{v}{\sqrt{2\pi}}=-\frac{1}{2}+\frac{y_v}{N}+\frac{1}{2N}$ for some integer $y_v$.  Using these arguments and the expression for $\mathbb{P}(\tilde{v}  |v)$ in Eq.~(\ref{eq:expressP}), we can bound for $v\in I$
\begin{eqnarray}
{\rm MSD}(\tilde{v})\leq \frac{2\pi}{N^4} \sum_{y=0}^{N-1} \frac{(y-y_v-\frac{1}{2})^2}{\sin^2(\frac{\pi}{N}(y-y_v-\frac{1}{2}))} \leq \nonumber \\
\frac{2\pi}{N^3}\max_{y-y_v\colon \frac{y_v}{N} \in [\alpha,1-\alpha]} \frac{(y-y_v-\frac{1}{2})^2}{\sin^2(\frac{\pi}{N}(y-y_v-\frac{1}{2}))}.
\label{eq:upper1}
\end{eqnarray}
Since the function $\frac{x^2}{\sin^2(x)}$ is even and monotonically increasing in $x$ for $x\geq 0$, the maximum is achieved for the largest possible $|y-y_v-\frac{1}{2}|$. We obtain
\begin{equation} 
{\rm MSD}(\tilde{v})\leq \frac{2\pi (1-\alpha)^2}{N \sin^2(\pi (\alpha-\frac{1}{2N}))}+\mathcal{O}\left(\frac{1}{N^2}\right). 
\label{eq:upper2}
\end{equation}
This shows that for finite $\alpha$, the MSD scales at most as $1/N=1/2^M$. In the limit when $\alpha \rightarrow 0$ one can upper bound the MSD by at most a constant, not decreasing with $N$, due to the fact that the wrong estimates for $v$ at the edges of the interval can have large error. 
The phase estimation protocol has a number of photons $\overline{n} \lessapprox \frac{\pi}{2} \sum_{k=0}^{M-1}2^{2k}=\frac{\pi}{6} (2^{2M}-1)=\frac{\pi}{6} (N^2-1)$. This holds since each round by some displacement $\alpha=2^{k-1} \sqrt{\pi}$ adds at most $\overline{n}=\frac{2^{2k}\pi}{4}$ photons, see Fig. \ref{fig:sequence}. Hence we upperbound ${\rm MSD}(\tilde{v}) = \mathcal{O}\left(\frac{1}{\sqrt{\overline{n}}}\right)$. This upperbound on the MSD is not necessarily tight, in particular the bounds in Eqs.~(\ref{eq:upper1}) and (\ref{eq:upper2}) can alter the dependence on $N=2^M \sim \sqrt{\overline{n}}$. It is an open question whether one can tighten these inequalities to explicitly prove that the MSD scales as $\frac{1}{\overline{n}}$.

As the phase estimation measurement for $S_p$ commutes with the phase estimation measurement for $S_q$, the probability $\mathbb{P}(\tilde{u} |u)$ is the same as for $\tilde{v}$ and $v$, and therefore the MSD is identical.

Due to having a finite mesh of values for $\tilde{v}$, the estimator $\tilde{v}$ (and similarly $\tilde{u}$), is not completely unbiased. However, inside the interval $I$, with a mesh of points with interdistance $\frac{1}{N}$, the bias $b(\tilde{v})=\mathcal{O}(\frac{1}{N})$ and thus ${\rm Var}(\tilde{v})={\rm MSD}(\tilde{v})-\mathcal{O}(\frac{1}{N^2})$, i.e. the difference between the mean squared deviation and the variance is negliglible in this context.

\subsubsection{Most Efficient Phase Estimation}
We make a side remark on what constitutes optimal phase estimation in this displacement setting. The best form of phase estimation requires one to optimize the accuracy on the phase, that is, minimizes the variance of $\theta$ in the phase $e^{i\theta}$, given an average number of photons $\overline{n}$ in the prepared state. A first remark is that since a variance or mean squared deviation performance metric does not give full information about the (shape) of the wavefunction of the grid state that one obtains, see comments in \cite{TW:GKP}, optimality may not only depend on the variance.
A second remark is that when $U$ is a displacement operator such as $S_p$ or $S_q$, performing a controlled-$U^k$ gate adds $\sim k^2$ photons to a state. Sequential phase estimation protocols which minimize the variance for a given overall running time $T$ have been proposed \cite{higgins+:heis} (with an improved analysis in \cite{KLY:PE}) under the assumption that applying $U^k$ takes time scaling {\em linearly} with $k$. Hence these protocols give {\em no advantage} in this setting as the number of photons scales quadratically, instead of linear with $k$ for (controlled)-$U^k$. 

\end{document}

%% file: displace_sensor16.bbl

%% file: displace_sensor16.bbl
\begin{thebibliography}{40}%
\makeatletter
\providecommand \@ifxundefined [1]{%
 \@ifx{#1\undefined}
}%
\providecommand \@ifnum [1]{%
 \ifnum #1\expandafter \@firstoftwo
 \else \expandafter \@secondoftwo
 \fi
}%
\providecommand \@ifx [1]{%
 \ifx #1\expandafter \@firstoftwo
 \else \expandafter \@secondoftwo
 \fi
}%
\providecommand \natexlab [1]{#1}%
\providecommand \enquote  [1]{``#1''}%
\providecommand \bibnamefont  [1]{#1}%
\providecommand \bibfnamefont [1]{#1}%
\providecommand \citenamefont [1]{#1}%
\providecommand \href@noop [0]{\@secondoftwo}%
\providecommand \href [0]{\begingroup \@sanitize@url \@href}%
\providecommand \@href[1]{\@@startlink{#1}\@@href}%
\providecommand \@@href[1]{\endgroup#1\@@endlink}%
\providecommand \@sanitize@url [0]{\catcode `\\12\catcode `\$12\catcode
  `\&12\catcode `\#12\catcode `\^12\catcode `\_12\catcode `\%12\relax}%
\providecommand \@@startlink[1]{}%
\providecommand \@@endlink[0]{}%
\providecommand \url  [0]{\begingroup\@sanitize@url \@url }%
\providecommand \@url [1]{\endgroup\@href {#1}{\urlprefix }}%
\providecommand \urlprefix  [0]{URL }%
\providecommand \Eprint [0]{\href }%
\providecommand \doibase [0]{http://dx.doi.org/}%
\providecommand \selectlanguage [0]{\@gobble}%
\providecommand \bibinfo  [0]{\@secondoftwo}%
\providecommand \bibfield  [0]{\@secondoftwo}%
\providecommand \translation [1]{[#1]}%
\providecommand \BibitemOpen [0]{}%
\providecommand \bibitemStop [0]{}%
\providecommand \bibitemNoStop [0]{.\EOS\space}%
\providecommand \EOS [0]{\spacefactor3000\relax}%
\providecommand \BibitemShut  [1]{\csname bibitem#1\endcsname}%
\let\auto@bib@innerbib\@empty
%</preamble>
\bibitem [{\citenamefont {Gottesman}\ \emph {et~al.}(2001)\citenamefont
  {Gottesman}, \citenamefont {Kitaev},\ and\ \citenamefont {Preskill}}]{GKP}%
  \BibitemOpen
  \bibfield  {author} {\bibinfo {author} {\bibfnamefont {D.}~\bibnamefont
  {Gottesman}}, \bibinfo {author} {\bibfnamefont {A.Yu.}\ \bibnamefont
  {Kitaev}}, \ and\ \bibinfo {author} {\bibfnamefont {J.}~\bibnamefont
  {Preskill}},\ }\bibfield  {title} {\enquote {\bibinfo {title} {Encoding a
  qubit in an oscillator},}\ }\href@noop {} {\bibfield  {journal} {\bibinfo
  {journal} {Phys. Rev. A}\ }\textbf {\bibinfo {volume} {64}},\ \bibinfo
  {pages} {012310} (\bibinfo {year} {2001})}\BibitemShut {NoStop}%
\bibitem [{\citenamefont {Terhal}\ and\ \citenamefont
  {Weigand}(2016)}]{TW:GKP}%
  \BibitemOpen
  \bibfield  {author} {\bibinfo {author} {\bibfnamefont {B.M.}\ \bibnamefont
  {Terhal}}\ and\ \bibinfo {author} {\bibfnamefont {D.}~\bibnamefont
  {Weigand}},\ }\bibfield  {title} {\enquote {\bibinfo {title} {Encoding a
  qubit into a cavity mode in circuit-{QED} using phase-estimation},}\
  }\href@noop {} {\bibfield  {journal} {\bibinfo  {journal} {Phys. Rev. A}\
  }\textbf {\bibinfo {volume} {93}},\ \bibinfo {pages} {012315} (\bibinfo
  {year} {2016})}\BibitemShut {NoStop}%
\bibitem [{Note1()}]{Note1}%
  \BibitemOpen
  \bibinfo {note} {Since we want to treat the field causing the displacement
  classically, it should contain a fair number of photons, but the effect on
  the oscillator should be weak, adding at most $\pi /2$ photons, hence one
  requires weak coupling and/or the oscillator and the signal being
  off-resonant.}\BibitemShut {Stop}%
\bibitem [{\citenamefont {{Regal}}\ \emph {et~al.}(2008)\citenamefont
  {{Regal}}, \citenamefont {{Teufel}},\ and\ \citenamefont
  {{Lehnert}}}]{RTL:nanomechanical}%
  \BibitemOpen
  \bibfield  {author} {\bibinfo {author} {\bibfnamefont {C.~A.}\ \bibnamefont
  {{Regal}}}, \bibinfo {author} {\bibfnamefont {J.~D.}\ \bibnamefont
  {{Teufel}}}, \ and\ \bibinfo {author} {\bibfnamefont {K.~W.}\ \bibnamefont
  {{Lehnert}}},\ }\bibfield  {title} {\enquote {\bibinfo {title} {{Measuring
  nanomechanical motion with a microwave cavity interferometer}},}\ }\href@noop
  {} {\bibfield  {journal} {\bibinfo  {journal} {ArXiv e-prints}\ } (\bibinfo
  {year} {2008})},\ \Eprint {http://arxiv.org/abs/0801.1827} {arXiv:0801.1827
  [quant-ph]} \BibitemShut {NoStop}%
\bibitem [{\citenamefont {Caves}\ \emph {et~al.}(1980)\citenamefont {Caves},
  \citenamefont {Thorne}, \citenamefont {Drever}, \citenamefont {Sandberg},\
  and\ \citenamefont {Zimmermann}}]{caves+:rmp}%
  \BibitemOpen
  \bibfield  {author} {\bibinfo {author} {\bibfnamefont {Carlton~M.}\
  \bibnamefont {Caves}}, \bibinfo {author} {\bibfnamefont {Kip~S.}\
  \bibnamefont {Thorne}}, \bibinfo {author} {\bibfnamefont {Ronald W.~P.}\
  \bibnamefont {Drever}}, \bibinfo {author} {\bibfnamefont {Vernon~D.}\
  \bibnamefont {Sandberg}}, \ and\ \bibinfo {author} {\bibfnamefont {Mark}\
  \bibnamefont {Zimmermann}},\ }\bibfield  {title} {\enquote {\bibinfo {title}
  {On the measurement of a weak classical force coupled to a quantum-mechanical
  oscillator. {I}. {I}ssues of principle},}\ }\href {\doibase
  10.1103/RevModPhys.52.341} {\bibfield  {journal} {\bibinfo  {journal} {Rev.
  Mod. Phys.}\ }\textbf {\bibinfo {volume} {52}},\ \bibinfo {pages} {341--392}
  (\bibinfo {year} {1980})}\BibitemShut {NoStop}%
\bibitem [{\citenamefont {{Giovannetti}}\ \emph {et~al.}(2004)\citenamefont
  {{Giovannetti}}, \citenamefont {{Lloyd}},\ and\ \citenamefont
  {{Maccone}}}]{GLM:quantum-enhanced}%
  \BibitemOpen
  \bibfield  {author} {\bibinfo {author} {\bibfnamefont {V.}~\bibnamefont
  {{Giovannetti}}}, \bibinfo {author} {\bibfnamefont {S.}~\bibnamefont
  {{Lloyd}}}, \ and\ \bibinfo {author} {\bibfnamefont {L.}~\bibnamefont
  {{Maccone}}},\ }\bibfield  {title} {\enquote {\bibinfo {title}
  {{Quantum-Enhanced Measurements: Beating the Standard Quantum Limit}},}\
  }\href {\doibase 10.1126/science.1104149} {\bibfield  {journal} {\bibinfo
  {journal} {Science}\ }\textbf {\bibinfo {volume} {306}},\ \bibinfo {pages}
  {1330--1336} (\bibinfo {year} {2004})},\ \Eprint
  {http://arxiv.org/abs/quant-ph/0412078} {quant-ph/0412078} \BibitemShut
  {NoStop}%
\bibitem [{\citenamefont {{Penasa}}\ \emph {et~al.}(2016)\citenamefont
  {{Penasa}}, \citenamefont {{Gerlich}}, \citenamefont {{Rybarczyk}},
  \citenamefont {{M{\'e}tillon}}, \citenamefont {{Brune}}, \citenamefont
  {{Raimond}}, \citenamefont {{Haroche}}, \citenamefont {{Davidovich}},\ and\
  \citenamefont {{Dotsenko}}}]{penasa+:displacement}%
  \BibitemOpen
  \bibfield  {author} {\bibinfo {author} {\bibfnamefont {M.}~\bibnamefont
  {{Penasa}}}, \bibinfo {author} {\bibfnamefont {S.}~\bibnamefont {{Gerlich}}},
  \bibinfo {author} {\bibfnamefont {T.}~\bibnamefont {{Rybarczyk}}}, \bibinfo
  {author} {\bibfnamefont {V.}~\bibnamefont {{M{\'e}tillon}}}, \bibinfo
  {author} {\bibfnamefont {M.}~\bibnamefont {{Brune}}}, \bibinfo {author}
  {\bibfnamefont {J.~M.}\ \bibnamefont {{Raimond}}}, \bibinfo {author}
  {\bibfnamefont {S.}~\bibnamefont {{Haroche}}}, \bibinfo {author}
  {\bibfnamefont {L.}~\bibnamefont {{Davidovich}}}, \ and\ \bibinfo {author}
  {\bibfnamefont {I.}~\bibnamefont {{Dotsenko}}},\ }\bibfield  {title}
  {\enquote {\bibinfo {title} {{Measurement of a microwave field amplitude
  beyond the standard quantum limit}},}\ }\href@noop {} {\bibfield  {journal}
  {\bibinfo  {journal} {ArXiv e-prints}\ } (\bibinfo {year} {2016})},\ \Eprint
  {http://arxiv.org/abs/1605.09568} {arXiv:1605.09568 [quant-ph]} \BibitemShut
  {NoStop}%
\bibitem [{\citenamefont {{\em et al}}(2016)}]{facon+:electrometer}%
  \BibitemOpen
  \bibfield  {author} {\bibinfo {author} {\bibfnamefont {Facon}\ \bibnamefont
  {{\em et al}}},\ }\bibfield  {title} {\enquote {\bibinfo {title} {{A
  sensitive electrometer based on a Rydberg atom in a Schroedinger-cat
  state}},}\ }\href@noop {} {\bibfield  {journal} {\bibinfo  {journal}
  {Nature}\ }\textbf {\bibinfo {volume} {535}},\ \bibinfo {pages} {262--265}
  (\bibinfo {year} {2016})}\BibitemShut {NoStop}%
\bibitem [{Note2()}]{Note2}%
  \BibitemOpen
  \bibinfo {note} {A (dispersive) phaseshift can come about when the linear
  coupling between oscillator and the external field is off-resonant while a
  real displacement is induced when the coupling is on-resonant.}\BibitemShut
  {Stop}%
\bibitem [{\citenamefont {Braunstein}\ and\ \citenamefont
  {Kimble}(2000)}]{BK:superdense}%
  \BibitemOpen
  \bibfield  {author} {\bibinfo {author} {\bibfnamefont {Samuel~L.}\
  \bibnamefont {Braunstein}}\ and\ \bibinfo {author} {\bibfnamefont {H.~J.}\
  \bibnamefont {Kimble}},\ }\bibfield  {title} {\enquote {\bibinfo {title}
  {Dense coding for continuous variables},}\ }\href {\doibase
  10.1103/PhysRevA.61.042302} {\bibfield  {journal} {\bibinfo  {journal} {Phys.
  Rev. A}\ }\textbf {\bibinfo {volume} {61}},\ \bibinfo {pages} {042302}
  (\bibinfo {year} {2000})}\BibitemShut {NoStop}%
\bibitem [{\citenamefont {{Flurin}}\ \emph {et~al.}(2012)\citenamefont
  {{Flurin}}, \citenamefont {{Roch}}, \citenamefont {{Mallet}}, \citenamefont
  {{Devoret}},\ and\ \citenamefont {{Huard}}}]{flurin+:2mode}%
  \BibitemOpen
  \bibfield  {author} {\bibinfo {author} {\bibfnamefont {E.}~\bibnamefont
  {{Flurin}}}, \bibinfo {author} {\bibfnamefont {N.}~\bibnamefont {{Roch}}},
  \bibinfo {author} {\bibfnamefont {F.}~\bibnamefont {{Mallet}}}, \bibinfo
  {author} {\bibfnamefont {M.~H.}\ \bibnamefont {{Devoret}}}, \ and\ \bibinfo
  {author} {\bibfnamefont {B.}~\bibnamefont {{Huard}}},\ }\bibfield  {title}
  {\enquote {\bibinfo {title} {{Generating Entangled Microwave Radiation Over
  Two Transmission Lines}},}\ }\href {\doibase 10.1103/PhysRevLett.109.183901}
  {\bibfield  {journal} {\bibinfo  {journal} {Physical Review Letters}\
  }\textbf {\bibinfo {volume} {109}},\ \bibinfo {eid} {183901} (\bibinfo {year}
  {2012})},\ \Eprint {http://arxiv.org/abs/1204.0732} {arXiv:1204.0732
  [cond-mat.mes-hall]} \BibitemShut {NoStop}%
\bibitem [{\citenamefont {{Mallet}}\ \emph {et~al.}(2011)\citenamefont
  {{Mallet}}, \citenamefont {{Castellanos-Beltran}}, \citenamefont {{Ku}},
  \citenamefont {{Glancy}}, \citenamefont {{Knill}}, \citenamefont {{Irwin}},
  \citenamefont {{Hilton}}, \citenamefont {{Vale}},\ and\ \citenamefont
  {{Lehnert}}}]{mallet+:squeezed}%
  \BibitemOpen
  \bibfield  {author} {\bibinfo {author} {\bibfnamefont {F.}~\bibnamefont
  {{Mallet}}}, \bibinfo {author} {\bibfnamefont {M.~A.}\ \bibnamefont
  {{Castellanos-Beltran}}}, \bibinfo {author} {\bibfnamefont {H.~S.}\
  \bibnamefont {{Ku}}}, \bibinfo {author} {\bibfnamefont {S.}~\bibnamefont
  {{Glancy}}}, \bibinfo {author} {\bibfnamefont {E.}~\bibnamefont {{Knill}}},
  \bibinfo {author} {\bibfnamefont {K.~D.}\ \bibnamefont {{Irwin}}}, \bibinfo
  {author} {\bibfnamefont {G.~C.}\ \bibnamefont {{Hilton}}}, \bibinfo {author}
  {\bibfnamefont {L.~R.}\ \bibnamefont {{Vale}}}, \ and\ \bibinfo {author}
  {\bibfnamefont {K.~W.}\ \bibnamefont {{Lehnert}}},\ }\bibfield  {title}
  {\enquote {\bibinfo {title} {{Quantum State Tomography of an Itinerant
  Squeezed Microwave Field}},}\ }\href {\doibase
  10.1103/PhysRevLett.106.220502} {\bibfield  {journal} {\bibinfo  {journal}
  {Physical Review Letters}\ }\textbf {\bibinfo {volume} {106}},\ \bibinfo
  {eid} {220502} (\bibinfo {year} {2011})},\ \Eprint
  {http://arxiv.org/abs/1012.0007} {arXiv:1012.0007 [quant-ph]} \BibitemShut
  {NoStop}%
\bibitem [{\citenamefont {Nakamura}\ and\ \citenamefont
  {Yamamoto}(2013)}]{NY:microwave}%
  \BibitemOpen
  \bibfield  {author} {\bibinfo {author} {\bibfnamefont {Y.}~\bibnamefont
  {Nakamura}}\ and\ \bibinfo {author} {\bibfnamefont {T.}~\bibnamefont
  {Yamamoto}},\ }\bibfield  {title} {\enquote {\bibinfo {title} {Breakthroughs
  in photonics 2012: Breakthroughs in microwave quantum photonics in
  superconducting circuits},}\ }\href {\doibase 10.1109/JPHOT.2013.2252005}
  {\bibfield  {journal} {\bibinfo  {journal} {IEEE Photonics Journal}\ }\textbf
  {\bibinfo {volume} {5}},\ \bibinfo {pages} {0701406--0701406} (\bibinfo
  {year} {2013})}\BibitemShut {NoStop}%
\bibitem [{\citenamefont {Zurek}(2001)}]{zurek:sub}%
  \BibitemOpen
  \bibfield  {author} {\bibinfo {author} {\bibfnamefont {W.}~\bibnamefont
  {Zurek}},\ }\bibfield  {title} {\enquote {\bibinfo {title} {Sub-{P}lanck
  structure in phase space and its relevance for quantum decoherence},}\
  }\href@noop {} {\bibfield  {journal} {\bibinfo  {journal} {Nature}\ }\textbf
  {\bibinfo {volume} {412}},\ \bibinfo {pages} {712--717} (\bibinfo {year}
  {2001})}\BibitemShut {NoStop}%
\bibitem [{\citenamefont {Vlastakis}\ \emph {et~al.}(2013)\citenamefont
  {Vlastakis}, \citenamefont {Kirchmair}, \citenamefont {Leghtas},
  \citenamefont {Nigg}, \citenamefont {Frunzio}, \citenamefont {Girvin},
  \citenamefont {Mirrahimi}, \citenamefont {Devoret},\ and\ \citenamefont
  {Schoelkopf}}]{vlastakis+:cat100}%
  \BibitemOpen
  \bibfield  {author} {\bibinfo {author} {\bibfnamefont {Brian}\ \bibnamefont
  {Vlastakis}}, \bibinfo {author} {\bibfnamefont {Gerhard}\ \bibnamefont
  {Kirchmair}}, \bibinfo {author} {\bibfnamefont {Zaki}\ \bibnamefont
  {Leghtas}}, \bibinfo {author} {\bibfnamefont {Simon~E.}\ \bibnamefont
  {Nigg}}, \bibinfo {author} {\bibfnamefont {Luigi}\ \bibnamefont {Frunzio}},
  \bibinfo {author} {\bibfnamefont {S.~M.}\ \bibnamefont {Girvin}}, \bibinfo
  {author} {\bibfnamefont {Mazyar}\ \bibnamefont {Mirrahimi}}, \bibinfo
  {author} {\bibfnamefont {M.~H.}\ \bibnamefont {Devoret}}, \ and\ \bibinfo
  {author} {\bibfnamefont {R.~J.}\ \bibnamefont {Schoelkopf}},\ }\bibfield
  {title} {\enquote {\bibinfo {title} {Deterministically encoding quantum
  information using 100-photon {S}chroedinger cat states},}\ }\href {\doibase
  10.1126/science.1243289} {\ \textbf {\bibinfo {volume} {342}},\ \bibinfo
  {pages} {607--610} (\bibinfo {year} {2013})}\BibitemShut {NoStop}%
\bibitem [{\citenamefont {Leghtas}\ \emph {et~al.}(2013)\citenamefont
  {Leghtas}, \citenamefont {Kirchmair}, \citenamefont {Vlastakis},
  \citenamefont {Schoelkopf}, \citenamefont {Devoret},\ and\ \citenamefont
  {Mirrahimi}}]{leghtas+:QEC}%
  \BibitemOpen
  \bibfield  {author} {\bibinfo {author} {\bibfnamefont {Z.}~\bibnamefont
  {Leghtas}}, \bibinfo {author} {\bibfnamefont {G.}~\bibnamefont {Kirchmair}},
  \bibinfo {author} {\bibfnamefont {B.}~\bibnamefont {Vlastakis}}, \bibinfo
  {author} {\bibfnamefont {R.J.}\ \bibnamefont {Schoelkopf}}, \bibinfo {author}
  {\bibfnamefont {M.~H.}\ \bibnamefont {Devoret}}, \ and\ \bibinfo {author}
  {\bibfnamefont {M.}~\bibnamefont {Mirrahimi}},\ }\bibfield  {title} {\enquote
  {\bibinfo {title} {Hardware-efficient autonomous quantum memory
  protection},}\ }\href@noop {} {\bibfield  {journal} {\bibinfo  {journal}
  {Phys. Rev. Lett.}\ }\textbf {\bibinfo {volume} {111}},\ \bibinfo {pages}
  {120501} (\bibinfo {year} {2013})}\BibitemShut {NoStop}%
\bibitem [{\citenamefont {Terhal}(2015)}]{BMT:review}%
  \BibitemOpen
  \bibfield  {author} {\bibinfo {author} {\bibfnamefont {Barbara~M.}\
  \bibnamefont {Terhal}},\ }\bibfield  {title} {\enquote {\bibinfo {title}
  {Quantum error correction for quantum memories},}\ }\href {\doibase
  10.1103/RevModPhys.87.307} {\bibfield  {journal} {\bibinfo  {journal} {Rev.
  Mod. Phys.}\ }\textbf {\bibinfo {volume} {87}},\ \bibinfo {pages} {307--346}
  (\bibinfo {year} {2015})}\BibitemShut {NoStop}%
\bibitem [{\citenamefont {Haroche}\ and\ \citenamefont
  {Raimond}(2006)}]{book:haroche}%
  \BibitemOpen
  \bibfield  {author} {\bibinfo {author} {\bibfnamefont {S.}~\bibnamefont
  {Haroche}}\ and\ \bibinfo {author} {\bibfnamefont {J.-M.}\ \bibnamefont
  {Raimond}},\ }\href@noop {} {\emph {\bibinfo {title} {Exploring the Quantum:
  Atoms, Cavities, and Photons}}}\ (\bibinfo  {publisher} {Oxford Univ.
  Press},\ \bibinfo {address} {Oxford},\ \bibinfo {year} {2006})\BibitemShut
  {NoStop}%
\bibitem [{\citenamefont {{Ofek}}\ \emph {et~al.}(2016)\citenamefont {{Ofek}},
  \citenamefont {{Petrenko}}, \citenamefont {{Heeres}}, \citenamefont
  {{Reinhold}}, \citenamefont {{Leghtas}}, \citenamefont {{Vlastakis}},
  \citenamefont {{Liu}}, \citenamefont {{Frunzio}}, \citenamefont {{Girvin}},
  \citenamefont {{Jiang}}, \citenamefont {{Mirrahimi}}, \citenamefont
  {{Devoret}},\ and\ \citenamefont {{Schoelkopf}}}]{ofek+:QEC}%
  \BibitemOpen
  \bibfield  {author} {\bibinfo {author} {\bibfnamefont {N.}~\bibnamefont
  {{Ofek}}}, \bibinfo {author} {\bibfnamefont {A.}~\bibnamefont {{Petrenko}}},
  \bibinfo {author} {\bibfnamefont {R.}~\bibnamefont {{Heeres}}}, \bibinfo
  {author} {\bibfnamefont {P.}~\bibnamefont {{Reinhold}}}, \bibinfo {author}
  {\bibfnamefont {Z.}~\bibnamefont {{Leghtas}}}, \bibinfo {author}
  {\bibfnamefont {B.}~\bibnamefont {{Vlastakis}}}, \bibinfo {author}
  {\bibfnamefont {Y.}~\bibnamefont {{Liu}}}, \bibinfo {author} {\bibfnamefont
  {L.}~\bibnamefont {{Frunzio}}}, \bibinfo {author} {\bibfnamefont {S.~M.}\
  \bibnamefont {{Girvin}}}, \bibinfo {author} {\bibfnamefont {L.}~\bibnamefont
  {{Jiang}}}, \bibinfo {author} {\bibfnamefont {M.}~\bibnamefont
  {{Mirrahimi}}}, \bibinfo {author} {\bibfnamefont {M.~H.}\ \bibnamefont
  {{Devoret}}}, \ and\ \bibinfo {author} {\bibfnamefont {R.~J.}\ \bibnamefont
  {{Schoelkopf}}},\ }\bibfield  {title} {\enquote {\bibinfo {title}
  {{Demonstrating Quantum Error Correction that Extends the Lifetime of Quantum
  Information}},}\ }\href@noop {} {\bibfield  {journal} {\bibinfo  {journal}
  {ArXiv e-prints}\ } (\bibinfo {year} {2016})},\ \Eprint
  {http://arxiv.org/abs/1602.04768} {arXiv:1602.04768 [quant-ph]} \BibitemShut
  {NoStop}%
\bibitem [{Note3()}]{Note3}%
  \BibitemOpen
  \bibinfo {note} {Equivalently, the displacement is $D(\beta )=\protect
  \qopname \relax o{exp}(\beta a^{\dagger }-\beta ^* a)$ with $u=\protect \sqrt
  {2} {\protect \rm Re}(\beta )$ and $v=\protect \sqrt {2} {\protect \rm
  Im}(\beta )$}\BibitemShut {NoStop}%
\bibitem [{\citenamefont {Helstrom}(1968)}]{helstrom:quantum_CR}%
  \BibitemOpen
  \bibfield  {author} {\bibinfo {author} {\bibfnamefont {C.W.}\ \bibnamefont
  {Helstrom}},\ }\bibfield  {title} {\enquote {\bibinfo {title} {The minimum
  variance of estimates in quantum signal detection},}\ }\href@noop {}
  {\bibfield  {journal} {\bibinfo  {journal} {{IEEE} Transactions on
  Information Theory}\ }\textbf {\bibinfo {volume} {14}},\ \bibinfo {pages}
  {234--242} (\bibinfo {year} {1968})}\BibitemShut {NoStop}%
\bibitem [{\citenamefont {Yuen}\ and\ \citenamefont {Lax}(1973)}]{yuen_lax}%
  \BibitemOpen
  \bibfield  {author} {\bibinfo {author} {\bibfnamefont {H.P.}\ \bibnamefont
  {Yuen}}\ and\ \bibinfo {author} {\bibfnamefont {M.}~\bibnamefont {Lax}},\
  }\bibfield  {title} {\enquote {\bibinfo {title} {Multiple-parameter quantum
  estimation and measurement of nonadjoint observables},}\ }\href@noop {}
  {\bibfield  {journal} {\bibinfo  {journal} {{IEEE} Transactions on
  Information Theory}\ }\textbf {\bibinfo {volume} {19}},\ \bibinfo {pages}
  {740--750} (\bibinfo {year} {1973})}\BibitemShut {NoStop}%
\bibitem [{\citenamefont {Braunstein}\ and\ \citenamefont
  {Caves}(1994)}]{BC:stat_dist}%
  \BibitemOpen
  \bibfield  {author} {\bibinfo {author} {\bibfnamefont {Samuel~L.}\
  \bibnamefont {Braunstein}}\ and\ \bibinfo {author} {\bibfnamefont
  {Carlton~M.}\ \bibnamefont {Caves}},\ }\bibfield  {title} {\enquote {\bibinfo
  {title} {Statistical distance and the geometry of quantum states},}\ }\href
  {\doibase 10.1103/PhysRevLett.72.3439} {\bibfield  {journal} {\bibinfo
  {journal} {Phys. Rev. Lett.}\ }\textbf {\bibinfo {volume} {72}},\ \bibinfo
  {pages} {3439--3443} (\bibinfo {year} {1994})}\BibitemShut {NoStop}%
\bibitem [{\citenamefont {{Paris}}(2009)}]{paris:LET}%
  \BibitemOpen
  \bibfield  {author} {\bibinfo {author} {\bibfnamefont {M.~G.~A.}\
  \bibnamefont {{Paris}}},\ }\bibfield  {title} {\enquote {\bibinfo {title}
  {{Quantum estimation for quantum technology}},}\ }\href@noop {} {\bibfield
  {journal} {\bibinfo  {journal} {Int.~J.~Quant.~Inf.~}\ }\textbf {\bibinfo
  {volume} {7}},\ \bibinfo {pages} {125} (\bibinfo {year} {2009})}\BibitemShut
  {NoStop}%
\bibitem [{\citenamefont {Genoni}\ \emph {et~al.}(2013)\citenamefont {Genoni},
  \citenamefont {Paris}, \citenamefont {Adesso}, \citenamefont {Nha},
  \citenamefont {Knight},\ and\ \citenamefont {Kim}}]{genoni+:displace}%
  \BibitemOpen
  \bibfield  {author} {\bibinfo {author} {\bibfnamefont {M.~G.}\ \bibnamefont
  {Genoni}}, \bibinfo {author} {\bibfnamefont {M.~G.~A.}\ \bibnamefont
  {Paris}}, \bibinfo {author} {\bibfnamefont {G.}~\bibnamefont {Adesso}},
  \bibinfo {author} {\bibfnamefont {H.}~\bibnamefont {Nha}}, \bibinfo {author}
  {\bibfnamefont {P.~L.}\ \bibnamefont {Knight}}, \ and\ \bibinfo {author}
  {\bibfnamefont {M.~S.}\ \bibnamefont {Kim}},\ }\bibfield  {title} {\enquote
  {\bibinfo {title} {Optimal estimation of joint parameters in phase space},}\
  }\href {\doibase 10.1103/PhysRevA.87.012107} {\bibfield  {journal} {\bibinfo
  {journal} {Phys. Rev. A}\ }\textbf {\bibinfo {volume} {87}},\ \bibinfo
  {pages} {012107} (\bibinfo {year} {2013})}\BibitemShut {NoStop}%
\bibitem [{\citenamefont {Wiseman}\ and\ \citenamefont
  {Milburn}(2010)}]{book:WM}%
  \BibitemOpen
  \bibfield  {author} {\bibinfo {author} {\bibfnamefont {H.}~\bibnamefont
  {Wiseman}}\ and\ \bibinfo {author} {\bibfnamefont {G.J.}\ \bibnamefont
  {Milburn}},\ }\href@noop {} {\emph {\bibinfo {title} {Quantum Measurement and
  Control}}}\ (\bibinfo  {publisher} {Cambridge University Press},\ \bibinfo
  {address} {Cambridge},\ \bibinfo {year} {2010})\BibitemShut {NoStop}%
\bibitem [{\citenamefont {{Szczykulska}}\ \emph {et~al.}(2016)\citenamefont
  {{Szczykulska}}, \citenamefont {{Baumgratz}},\ and\ \citenamefont
  {{Datta}}}]{SBD:multi-parameter}%
  \BibitemOpen
  \bibfield  {author} {\bibinfo {author} {\bibfnamefont {M.}~\bibnamefont
  {{Szczykulska}}}, \bibinfo {author} {\bibfnamefont {T.}~\bibnamefont
  {{Baumgratz}}}, \ and\ \bibinfo {author} {\bibfnamefont {A.}~\bibnamefont
  {{Datta}}},\ }\bibfield  {title} {\enquote {\bibinfo {title}
  {{Multi-parameter Quantum Metrology}},}\ }\href@noop {} {\bibfield  {journal}
  {\bibinfo  {journal} {ArXiv e-prints}\ } (\bibinfo {year} {2016})},\ \Eprint
  {http://arxiv.org/abs/1604.02615} {arXiv:1604.02615 [quant-ph]} \BibitemShut
  {NoStop}%
\bibitem [{\citenamefont {{Boixo}}\ \emph {et~al.}(2007)\citenamefont
  {{Boixo}}, \citenamefont {{Flammia}}, \citenamefont {{Caves}},\ and\
  \citenamefont {{Geremia}}}]{boixo+:metro}%
  \BibitemOpen
  \bibfield  {author} {\bibinfo {author} {\bibfnamefont {S.}~\bibnamefont
  {{Boixo}}}, \bibinfo {author} {\bibfnamefont {S.~T.}\ \bibnamefont
  {{Flammia}}}, \bibinfo {author} {\bibfnamefont {C.~M.}\ \bibnamefont
  {{Caves}}}, \ and\ \bibinfo {author} {\bibfnamefont {J.}~\bibnamefont
  {{Geremia}}},\ }\bibfield  {title} {\enquote {\bibinfo {title} {{Generalized
  Limits for Single-Parameter Quantum Estimation}},}\ }\href {\doibase
  10.1103/PhysRevLett.98.090401} {\bibfield  {journal} {\bibinfo  {journal}
  {Physical Review Letters}\ }\textbf {\bibinfo {volume} {98}},\ \bibinfo {eid}
  {090401} (\bibinfo {year} {2007})},\ \Eprint
  {http://arxiv.org/abs/quant-ph/0609179} {quant-ph/0609179} \BibitemShut
  {NoStop}%
\bibitem [{Note4()}]{Note4}%
  \BibitemOpen
  \bibinfo {note} {For a displaced photon number state sensor $D(\beta )
  \mathinner {|{n}\delimiter "526930B }$, a measurement in the photon number
  basis would not allow one to resolve small displacements $|\beta |^2 \ll 1$.
  A measurement in the overcomplete basis $\mathinner {|{\psi _{\protect
  \mathaccentV {tilde}07E{\beta }}}\delimiter "526930B }=D(\protect
  \mathaccentV {tilde}07E{\beta })\mathinner {|{n}\delimiter "526930B }$ would
  output the estimate $\protect \mathaccentV {tilde}07E{\beta }$ with
  probability $\protect \mathbb {P}(\protect \mathaccentV {tilde}07E{\beta
  }|\beta )=\protect \frac {1}{\pi } |\mathinner {\delimiter "426830A {n}|}
  D(\beta -\protect \mathaccentV {tilde}07E{\beta }) \mathinner {|{n}\delimiter
  "526930B }|^2=\protect \frac {1}{\pi } e^{-|\beta -\protect \mathaccentV
  {tilde}07E{\beta }|^2}(L_n(|\beta -\protect \mathaccentV {tilde}07E{\beta
  }|^2))^2$ with Laguerre polynomial $L_n(x)$, whose support for $x > 0$
  increases for larger $n$.}\BibitemShut {Stop}%
\bibitem [{\citenamefont {Wehrl}(1979)}]{wehrl:1979}%
  \BibitemOpen
  \bibfield  {author} {\bibinfo {author} {\bibfnamefont {Alfred}\ \bibnamefont
  {Wehrl}},\ }\bibfield  {title} {\enquote {\bibinfo {title} {On the relation
  between classical and quantum-mechanical entropy},}\ }\href {\doibase
  http://dx.doi.org/10.1016/0034-4877(79)90070-3} {\bibfield  {journal}
  {\bibinfo  {journal} {Reports on Mathematical Physics}\ }\textbf {\bibinfo
  {volume} {16}},\ \bibinfo {pages} {353 -- 358} (\bibinfo {year}
  {1979})}\BibitemShut {NoStop}%
\bibitem [{\citenamefont {{Sun}}\ \emph {et~al.}(2014)\citenamefont {{Sun}},
  \citenamefont {{Petrenko}}, \citenamefont {{Leghtas}}, \citenamefont
  {{Vlastakis}}, \citenamefont {{Kirchmair}}, \citenamefont {{Sliwa}},
  \citenamefont {{Narla}}, \citenamefont {{Hatridge}}, \citenamefont
  {{Shankar}}, \citenamefont {{Blumoff}}, \citenamefont {{Frunzio}},
  \citenamefont {{Mirrahimi}}, \citenamefont {{Devoret}},\ and\ \citenamefont
  {{Schoelkopf}}}]{sun+:parity}%
  \BibitemOpen
  \bibfield  {author} {\bibinfo {author} {\bibfnamefont {L.}~\bibnamefont
  {{Sun}}}, \bibinfo {author} {\bibfnamefont {A.}~\bibnamefont {{Petrenko}}},
  \bibinfo {author} {\bibfnamefont {Z.}~\bibnamefont {{Leghtas}}}, \bibinfo
  {author} {\bibfnamefont {B.}~\bibnamefont {{Vlastakis}}}, \bibinfo {author}
  {\bibfnamefont {G.}~\bibnamefont {{Kirchmair}}}, \bibinfo {author}
  {\bibfnamefont {K.~M.}\ \bibnamefont {{Sliwa}}}, \bibinfo {author}
  {\bibfnamefont {A.}~\bibnamefont {{Narla}}}, \bibinfo {author} {\bibfnamefont
  {M.}~\bibnamefont {{Hatridge}}}, \bibinfo {author} {\bibfnamefont
  {S.}~\bibnamefont {{Shankar}}}, \bibinfo {author} {\bibfnamefont
  {J.}~\bibnamefont {{Blumoff}}}, \bibinfo {author} {\bibfnamefont
  {L.}~\bibnamefont {{Frunzio}}}, \bibinfo {author} {\bibfnamefont
  {M.}~\bibnamefont {{Mirrahimi}}}, \bibinfo {author} {\bibfnamefont {M.~H.}\
  \bibnamefont {{Devoret}}}, \ and\ \bibinfo {author} {\bibfnamefont {R.~J.}\
  \bibnamefont {{Schoelkopf}}},\ }\bibfield  {title} {\enquote {\bibinfo
  {title} {{Tracking photon jumps with repeated quantum non-demolition parity
  measurements}},}\ }\href {\doibase 10.1038/nature13436} {\bibfield  {journal}
  {\bibinfo  {journal} {\nat}\ }\textbf {\bibinfo {volume} {511}},\ \bibinfo
  {pages} {444--448} (\bibinfo {year} {2014})},\ \Eprint
  {http://arxiv.org/abs/1311.2534} {arXiv:1311.2534 [quant-ph]} \BibitemShut
  {NoStop}%
\bibitem [{\citenamefont {{Heeres}}\ \emph {et~al.}(2016)\citenamefont
  {{Heeres}}, \citenamefont {{Reinhold}}, \citenamefont {{Ofek}}, \citenamefont
  {{Frunzio}}, \citenamefont {{Jiang}}, \citenamefont {{Devoret}},\ and\
  \citenamefont {{Schoelkopf}}}]{heeres+:control}%
  \BibitemOpen
  \bibfield  {author} {\bibinfo {author} {\bibfnamefont {R.~W.}\ \bibnamefont
  {{Heeres}}}, \bibinfo {author} {\bibfnamefont {P.}~\bibnamefont
  {{Reinhold}}}, \bibinfo {author} {\bibfnamefont {N.}~\bibnamefont {{Ofek}}},
  \bibinfo {author} {\bibfnamefont {L.}~\bibnamefont {{Frunzio}}}, \bibinfo
  {author} {\bibfnamefont {L.}~\bibnamefont {{Jiang}}}, \bibinfo {author}
  {\bibfnamefont {M.~H.}\ \bibnamefont {{Devoret}}}, \ and\ \bibinfo {author}
  {\bibfnamefont {R.~J.}\ \bibnamefont {{Schoelkopf}}},\ }\bibfield  {title}
  {\enquote {\bibinfo {title} {{Implementing a Universal Gate Set on a Logical
  Qubit Encoded in an Oscillator}},}\ }\href@noop {} {\bibfield  {journal}
  {\bibinfo  {journal} {ArXiv e-prints}\ } (\bibinfo {year} {2016})},\ \Eprint
  {http://arxiv.org/abs/1608.02430} {arXiv:1608.02430 [quant-ph]} \BibitemShut
  {NoStop}%
\bibitem [{\citenamefont {{Bruno}}\ \emph {et~al.}(2015)\citenamefont
  {{Bruno}}, \citenamefont {{de Lange}}, \citenamefont {{Asaad}}, \citenamefont
  {{van der Enden}}, \citenamefont {{Langford}},\ and\ \citenamefont
  {{DiCarlo}}}]{BDD15}%
  \BibitemOpen
  \bibfield  {author} {\bibinfo {author} {\bibfnamefont {A.}~\bibnamefont
  {{Bruno}}}, \bibinfo {author} {\bibfnamefont {G.}~\bibnamefont {{de Lange}}},
  \bibinfo {author} {\bibfnamefont {S.}~\bibnamefont {{Asaad}}}, \bibinfo
  {author} {\bibfnamefont {K.~L.}\ \bibnamefont {{van der Enden}}}, \bibinfo
  {author} {\bibfnamefont {N.~K.}\ \bibnamefont {{Langford}}}, \ and\ \bibinfo
  {author} {\bibfnamefont {L.}~\bibnamefont {{DiCarlo}}},\ }\bibfield  {title}
  {\enquote {\bibinfo {title} {{Reducing intrinsic loss in superconducting
  resonators by surface treatment and deep etching of silicon substrates}},}\
  }\href@noop {} {\bibfield  {journal} {\bibinfo  {journal} {ArXiv e-prints}\ }
  (\bibinfo {year} {2015})},\ \Eprint {http://arxiv.org/abs/1502.04082}
  {arXiv:1502.04082 [cond-mat.supr-con]} \BibitemShut {NoStop}%
\bibitem [{\citenamefont {{Braunstein}}\ \emph {et~al.}(1996)\citenamefont
  {{Braunstein}}, \citenamefont {{Caves}},\ and\ \citenamefont
  {{Milburn}}}]{BCM:uncertainty}%
  \BibitemOpen
  \bibfield  {author} {\bibinfo {author} {\bibfnamefont {S.}~\bibnamefont
  {{Braunstein}}}, \bibinfo {author} {\bibfnamefont {C.}~\bibnamefont
  {{Caves}}}, \ and\ \bibinfo {author} {\bibfnamefont {G}~\bibnamefont
  {{Milburn}}},\ }\bibfield  {title} {\enquote {\bibinfo {title} {{Generalized
  Uncertainty Relations: Theory, Examples, and Lorentz Invariance}},}\
  }\href@noop {} {\bibfield  {journal} {\bibinfo  {journal} {Annals of
  Physics}\ }\textbf {\bibinfo {volume} {247}},\ \bibinfo {pages} {135--173}
  (\bibinfo {year} {1996})}\BibitemShut {NoStop}%
\bibitem [{\citenamefont {Mosca}\ and\ \citenamefont
  {Ekert}(1998)}]{ME:hidden}%
  \BibitemOpen
  \bibfield  {author} {\bibinfo {author} {\bibfnamefont {M.}~\bibnamefont
  {Mosca}}\ and\ \bibinfo {author} {\bibfnamefont {A.}~\bibnamefont {Ekert}},\
  }\bibfield  {title} {\enquote {\bibinfo {title} {The hidden subgroup problem
  and eigenvalue estimation on a quantum computer},}\ }in\ \href@noop {} {\emph
  {\bibinfo {booktitle} {Proceedings of 1st NASA QCQC conference}}},\ \bibinfo
  {series} {Lecture Notes in Computer Science}, Vol.\ \bibinfo {volume} {1509}\
  (\bibinfo  {publisher} {Springer},\ \bibinfo {year} {1998})\BibitemShut
  {NoStop}%
\bibitem [{Note5()}]{Note5}%
  \BibitemOpen
  \bibinfo {note} {If we pre-displace the bosonic mode by $S_p^{-1/2}$ in the
  $k=0$ circuit for $S_p$, see Fig.~\ref {fig:SPE_part} (and similarly for
  $S_q$), the $S_p$ and $S_q$ circuits would not commute, but the commutation
  would simply induce an additional rotation on a qubit since the gate
  $S_p^{-1/2}$ {\protect \em does not commute} with the qubit-controlled $S_q$
  gate and similarly for the circuit of phase estimation for
  $S_q$.}\BibitemShut {Stop}%
\bibitem [{\citenamefont {Nielsen}\ and\ \citenamefont
  {Chuang}(2000)}]{book:nielsen&chuang}%
  \BibitemOpen
  \bibfield  {author} {\bibinfo {author} {\bibfnamefont {M.~A.}\ \bibnamefont
  {Nielsen}}\ and\ \bibinfo {author} {\bibfnamefont {I.~L.}\ \bibnamefont
  {Chuang}},\ }\href@noop {} {\emph {\bibinfo {title} {Quantum Computation and
  Quantum Information}}}\ (\bibinfo  {publisher} {Cambridge University Press},\
  \bibinfo {address} {Cambridge, U.K.},\ \bibinfo {year} {2000})\BibitemShut
  {NoStop}%
\bibitem [{\citenamefont {Cleve}\ \emph {et~al.}(1998)\citenamefont {Cleve},
  \citenamefont {Ekert}, \citenamefont {Macchiavello},\ and\ \citenamefont
  {Mosca}}]{cleve+:revisited}%
  \BibitemOpen
  \bibfield  {author} {\bibinfo {author} {\bibfnamefont {R.}~\bibnamefont
  {Cleve}}, \bibinfo {author} {\bibfnamefont {A.}~\bibnamefont {Ekert}},
  \bibinfo {author} {\bibfnamefont {C.}~\bibnamefont {Macchiavello}}, \ and\
  \bibinfo {author} {\bibfnamefont {M.}~\bibnamefont {Mosca}},\ }\bibfield
  {title} {\enquote {\bibinfo {title} {Quantum algorithms revisited},}\ }in\
  \href@noop {} {\emph {\bibinfo {booktitle} {Proceedings of the Royal Society
  of London}}},\ Vol.\ \bibinfo {volume} {A454}\ (\bibinfo {year} {1998})\ pp.\
  \bibinfo {pages} {339--354}\BibitemShut {NoStop}%
\bibitem [{\citenamefont {{Higgins}}\ \emph {et~al.}(2009)\citenamefont
  {{Higgins}}, \citenamefont {{Berry}}, \citenamefont {{Bartlett}},
  \citenamefont {{Mitchell}}, \citenamefont {{Wiseman}},\ and\ \citenamefont
  {{Pryde}}}]{higgins+:heis}%
  \BibitemOpen
  \bibfield  {author} {\bibinfo {author} {\bibfnamefont {B.~L.}\ \bibnamefont
  {{Higgins}}}, \bibinfo {author} {\bibfnamefont {D.~W.}\ \bibnamefont
  {{Berry}}}, \bibinfo {author} {\bibfnamefont {S.~D.}\ \bibnamefont
  {{Bartlett}}}, \bibinfo {author} {\bibfnamefont {M.~W.}\ \bibnamefont
  {{Mitchell}}}, \bibinfo {author} {\bibfnamefont {H.~M.}\ \bibnamefont
  {{Wiseman}}}, \ and\ \bibinfo {author} {\bibfnamefont {G.~J.}\ \bibnamefont
  {{Pryde}}},\ }\bibfield  {title} {\enquote {\bibinfo {title} {{Demonstrating
  Heisenberg-limited unambiguous phase estimation without adaptive
  measurements}},}\ }\href {\doibase 10.1088/1367-2630/11/7/073023} {\bibfield
  {journal} {\bibinfo  {journal} {New Journal of Physics}\ }\textbf {\bibinfo
  {volume} {11}},\ \bibinfo {eid} {073023} (\bibinfo {year} {2009})},\ \Eprint
  {http://arxiv.org/abs/0809.3308} {arXiv:0809.3308 [quant-ph]} \BibitemShut
  {NoStop}%
\bibitem [{\citenamefont {Kimmel}\ \emph {et~al.}(2015)\citenamefont {Kimmel},
  \citenamefont {Low},\ and\ \citenamefont {Yoder}}]{KLY:PE}%
  \BibitemOpen
  \bibfield  {author} {\bibinfo {author} {\bibfnamefont {Shelby}\ \bibnamefont
  {Kimmel}}, \bibinfo {author} {\bibfnamefont {Guang~Hao}\ \bibnamefont {Low}},
  \ and\ \bibinfo {author} {\bibfnamefont {Theodore~J.}\ \bibnamefont
  {Yoder}},\ }\bibfield  {title} {\enquote {\bibinfo {title} {Robust
  calibration of a universal single-qubit gate set via robust phase
  estimation},}\ }\href {\doibase 10.1103/PhysRevA.92.062315} {\bibfield
  {journal} {\bibinfo  {journal} {Phys. Rev. A}\ }\textbf {\bibinfo {volume}
  {92}},\ \bibinfo {pages} {062315} (\bibinfo {year} {2015})}\BibitemShut
  {NoStop}%
\end{thebibliography}
